\documentclass[11pt,a4paper]{article}
\usepackage{jcappub}

\usepackage{amsmath,amssymb,bm,float,mathrsfs,subfigure,tabularx}
\usepackage{graphicx}
\usepackage{dcolumn}
\usepackage{color}
\usepackage{hyperref}

  \def\vs#1{\vspace{#1\baselineskip}}
  
  \newcommand{\beq}{\begin{equation}}
  \newcommand{\eeq}{\end{equation}}
  \newcommand{\al}[1]{\begin{align} #1 \end{align}}
  \newcommand{\bi}{\begin{itemize}}
  \newcommand{\ei}{\end{itemize}}
  \newcommand{\bc}{\begin{center}}
  \newcommand{\ec}{\end{center}}
  \newcommand{\rom}[1]{_{\mathrm{#1}}}
  \def\dd{\mathrm{d}}
  
  \def\rmg{\mathrm{g}}
  \def\rmm{\mathrm{m}}
  \def\rmB{\mathrm{B}}
  
  \def\rmE{\mathrm{E}}

  \def\rmO{\mathrm{O}}
  \def\rmS{\mathrm{S}}

  \def\rmX{\mathrm{X}}
  
  \def\mcB{\mathcal{B}}
  
  \def\mcD{\mathcal{D}}
  \def\mcE{\mathcal{E}}

  \def\mcN{\mathcal{N}}
  \def\mcO{\mathcal{O}}

  \def\mcS{\mathcal{S}}
  \def\mcT{\mathcal{T}}
  
  \def\mcV{\mathcal{V}}

  \def\pd{\partial}
  \def\Tr{\mbox{Tr}}

  \def\pspin{\ooalign{\hfil/\hfil\crcr$\partial$}}
  \def\mspin{\bar{\ooalign{\hfil/\hfil\crcr$\partial$}}}
  \newcommand{\ave}[1]{\left\langle #1 \right\rangle}

  \def\3Dint#1{\int\frac{\dd^{3}{#1 }}{(2\pi )^3}}

  \def\hatn{\hat{\bm n}}
  \def\hatk{\hat{\bm k}}
  
  \def\lO{\lambda_{\rm O}}
  \def\cS{\chi_{\rm S}}

\title{Weak lensing generated by vector perturbations 
and 
detectability of cosmic strings}

\author[a]{Daisuke Yamauchi}
\author[b]{Toshiya Namikawa}
\author[c,d]{Atsushi Taruya}

\affiliation[a]{%
Institute for Cosmic Ray Research, The University of Tokyo, Kashiwa, Chiba 277-8582, Japan
}%
\affiliation[b]{%
Department of Physics, Graduate School of Science, The University of Tokyo, Bunkyo-ku, Tokyo 113-033, Japan
}%
\affiliation[c]{%
Research Center for the Early Universe, Graduate School of Science, The University of Tokyo, Bunkyo-ku, Tokyo 113-033, Japan
}%
\affiliation[d]{%
Kavli Institute for the Physics and Mathematics of the Universe, The University 
of Tokyo, Kashiwa, Chiba 277-8568, Japan 
}%

\emailAdd{yamauchi@icrr.u-tokyo.ac.jp}
\emailAdd{namikawa@utap.phys.s.u-tokyo.ac.jp}
\emailAdd{ataruya@utap.phys.s.u-tokyo.ac.jp}

\abstract{
We study the observational signature of 
vector metric perturbations through the effect of weak gravitational lensing.  
In the presence of vector perturbations, 
the non-vanishing 
signals for B-mode cosmic shear and curl-mode deflection angle, 
which have never appeared in the case of scalar metric perturbations, 
naturally arise. Solving the geodesic and geodesic deviation equations, 
we drive the full-sky formulas for angular power spectra of weak lensing 
signals, and give the explicit expressions for E-/B-mode cosmic shear and 
gradient-/curl-mode deflection angle. 
As a possible source for seeding vector perturbations, 
we then consider a cosmic string network, and discuss 
its detectability from upcoming weak lensing and CMB measurements. 
Based on the formulas 
and a simple model for cosmic string network, 
we calculate the angular power spectra and expected signal-to-noise ratios 
for the B-mode cosmic shear and curl-mode deflection angle. We find that  
the weak lensing signals are enhanced for a smaller intercommuting probability 
of the string network, $P$, and they are potentially detectable from 
the upcoming cosmic shear and CMB lensing observations. For $P\sim 10^{-1}$, 
the minimum detectable tension of the cosmic string will be down to 
$G\mu\sim 5\times 10^{-8}$. With a theoretically inferred smallest value 
$P\sim 10^{-3}$, we could even detect the string with 
$G\mu\sim 5\times 10^{-10}$. 
}

\begin{document}

\maketitle
\tableofcontents

\section{Introduction} \label{sec:introduction} 

In standard cosmology, the vector mode of metric perturbations is thought 
to be a very minor component, and it does not serve as a seed of 
structure formation. 
One of the main theoretical reasons why we usually neglect vector perturbation 
is that in the absence of sources, vector perturbations 
decay away, and rapidly become negligible as the universe expands. 
It is, however, known that vector perturbations are generated 
via a variety of mechanisms in the early universe. 
Possible sources to generate vector perturbations include 
topological defects such as cosmic strings~\cite{Battye:2010xz,Bevis:2010gj,Bevis:2007qz,Pogosian:2007gi,Seljak:2006hi,Wyman:2005tu,Pogosian:2003mz},
anisotropic stress of magnetic field~\cite{Subramanian:2003sh,Lewis:2004ef,Lewis:2004kg,Paoletti:2008ck,Shaw:2009nf}, 
massive neutrinos~\cite{Lewis:2004kg,Shaw:2009nf},
second-order primordial density perturbations~\cite{Ichiki:2006cd,Lu:2007cj,Lu:2008ju,Christopherson:2009bt,Christopherson:2010ek,Christopherson:2010dw}, 
and modification of vector sector of gravity such as Einstein-Aether 
theory~\cite{Lim:2004js,Zuntz:2008zz,Zuntz:2010jp,ArmendarizPicon:2010rs,Nakashima:2011fu}. In particular, there are active mechanisms that 
continuously generate vector perturbations even at late-time epoch. One such 
example is those produced by topological defects. Hence, even with a 
tiny fraction, active seeds can induce the non-vanishing signals of 
vector perturbations at present time, which might be potentially 
detectable through precision cosmological observations. 
A search for those tiny signals is thus very interesting and valuable, and 
the detection and/or measurement of vector perturbation offers 
an important clue to probe the physics and history of the very 
early universe beyond the last scattering surface.

In this paper, among various cosmological observations, 
we are particularly interested in the weak lensing observations, which 
can provide a direct evidence for the intervening vector perturbations along 
a line of sight by measuring the spatial patterns on the deformation of 
photon path. 
The weak lensing measurements of background sources such as galaxies and 
cosmic microwave background (CMB) have been widely studied and now been 
accepted as a standard cosmological technique 
~\cite{Bernardeau:2011tc,Bernardeau:2009bm,Seitz:1994xf,Kaiser:1996tp,
Blandford:1991zz,Sasaki:1987ad,1961RSPSA.264..309S,Hu:2000ee,Stebbins:1996wx,Kamionkowski:1997mp} 
(for reviews, see \cite{Lewis:2006fu,Perlick:2004tq,Bartelmann:1999yn,
Sasaki:1993tu}). There are a number of planned wide and deep weak lensing 
surveys, including 
Subaru Hyper Suprime-Cam (HSC) survey~\cite{HSC}, Dark Energy Survey 
(DES)~\cite{astro-ph/0510346}, and Large Synaptic Survey 
Telescope (LSST)~\cite{arXiv:0912.0201}. They will 
provide a high-precision measurement of the deformation of the 
distant-galaxy images, namely cosmic shear fields. On the other hand, 
ongoing and upcoming CMB experiments such as PLANCK~\cite{:2006uk}, 
POLARBEAR~\cite{arXiv:1011.0763}, ACTPol~\cite{arXiv:1006.5049}, 
SPTPol~\cite{2009AIPC.1185..511M},  
CMBPol~\cite{Baumann:2008aq}, and COrE~\cite{arXiv:1102.2181}, 
offer a unique opportunity to probe the gravitational lensing 
deflection of the CMB photons, called CMB-lensing signals, with 
unprecedented precision.

With the increasing interest in the precision weak lensing measurements, 
in this paper, we intend to clarify the observational signature of vector 
perturbations on the weak lensing experiment. 
The spatial pattern of cosmic shear fields 
is generally described by a two-dimensional symmetric trace-free field 
on the sky, and it 
can be decomposed into two parts; even-parity mode (E-mode)
and the odd-parity mode (B-mode) (e.g., \cite{Stebbins:1996wx,Kamionkowski:1997mp}). 
Similarly, 
the deflection angle is decomposed into a gradient of scalar lensing
potential (gradient-mode) and a rotation of pseudo-scalar lensing potential
(curl-mode) (e.g., \cite{Stebbins:1996wx,Hirata:2003ka}). 
The symmetric argument implies that the B-mode shear 
and the curl-mode deflection angle 
are produced by the vector and tensor perturbations, 
but not by the scalar perturbations. 
Hence, the non-vanishing B-mode or curl-mode signal on large angular 
scales would be a direct evidence for non-scalar metric
perturbations. The weak lensing effect by the tensor perturbations has 
been previously studied in the cases of primordial gravitational wave (GW) 
\cite{Dodelson:2003bv,Cooray:2005hm,Li:2006si} and secondary GW generated 
by the second-order primordial density perturbations~\cite{Sarkar:2008ii}, 
but the effect turns out to be very small and difficult to observe 
(but see Refs.~\cite{Book:2011dz,Masui:2010cz}).
Here, we consider the weak lensing generated by vector perturbations, and  
derive the useful formulas for angular power spectra of 
E-/B-mode cosmic shear 
[eqs.~\eqref{eq:shear EE BB power spectrum}, \eqref{eq:E transfer function}
--\eqref{eq:beta^1 def}],  
and the gradient-/curl-mode deflection angle 
[eqs.~\eqref{eq:scalar and pseudo-scalar power spectrum},
\eqref{eq:phi transfer function}, and \eqref{eq:varpi transfer function}].
As a prospect for detecting non-zero B-mode cosmic shear or curl-mode 
deflection angle, we consider a cosmic string network as a possible source 
for seeding vector perturbations, and discuss its detectability.

It is known that 
the cosmic strings might have emerged in the early universe through spontaneous 
symmetry breakings~\cite{Kibble:1976sj,Hindmarsh:1994re,Perivolaropoulos:2005wa,
Jeannerot:2003qv}. Recently, another possibility to produce cosmic string 
has been pointed out in the context of superstring theory, and it is called 
cosmic superstring. The properties of cosmic superstrings are 
quite similar to those of ordinary 
cosmic strings~\cite{Sarangi:2002yt,Jones:2003da,Copeland:2003bj,
Dvali:2003zj,Kachru:2003sx}  (for reviews, see \cite{Polchinski:2004ia,
Polchinski:2004hb,Davis:2005dd,Copeland:2009ga,Sakellariadou:2009ev,
Ringeval:2010ca,Majumdar:2005qc,Copeland:2011dx}), except for the fact that 
the intercommuting probability between strings is relatively low. Thus, 
not only the string tension, $\mu$, but also the intercommuting probability, 
$P$,  are the important parameters to characterize the dynamics of cosmic 
string, as well as 
to distinguish between the conventional cosmic strings and the 
cosmic superstrings~\cite{Namikawa:2011cs,Takahashi:2008ui,Yamauchi:2010vy,Yamauchi:2010ms,Yamauchi:2011cu}. 
Currently, the tightest observational constraint on 
$\mu$ and $P$ are obtained from CMB observations through 
Gott-Kaiser-Stebbins (GKS) effect~\cite{Kaiser:1984iv,Gott:1984ef,Hindmarsh:1993pu}, which is basically imprinted on small angular 
scales~\cite{Battye:2010xz,Bevis:2010gj,Bevis:2007qz,Pogosian:2007gi,Seljak:2006hi,Wyman:2005tu,Pogosian:2003mz,Yamauchi:2010ms,Dunkley:2010ge}. 
Theoretically, the parameter $P$ is expected to 
lies in $10^{-3}\lesssim P\lesssim 1$~\cite{Polchinski:1988cn,Jackson:2004zg,Hanany:2005bc} (though the range of parameters strongly depends on the type of 
strings and the detail of the model), and the current observation is 
not enough to constrain a wide parameter range of $P$. 
In this paper, based on the formula for weak lensing power spectra and a simple
model of cosmic string network, we calculate the power spectra of B-mode 
cosmic shear and curl-mode deflection angle.  The possibility to detect the 
weak lensing signals from the vector perturbations is 
discussed in detail for specific weak lensing and CMB measurements
(see also \cite{Benabed:2000tr,Bernardeau:2000xu,Uzan:2000xv,Thomas:2009bm,Yamauchi:2011cu}).

The paper is organized as follows. 
In section \ref{sec:Basic equations for weak lensing},
we give basic equations for the weak lensing, and derive the expression 
for the deflection angle and the Jacobi map induced by the vector perturbations.
In section \ref{sec:Weak lensing observations induced by vector perturbations},
we investigate the properties of the shear fields and the deflection angle, 
and derive the formulas of the angular power spectra for the E-/B-mode 
cosmic shear and the gradient-/curl-mode deflection angle.
Based on the formulas, 
in section \ref{sec:Implications to a cosmic string network}, 
prospects for measuring the B-mode cosmic shear or 
curl-mode deflection angle are discussed, especially focusing on the cosmic 
string network.  
Finally, section \ref{sec:Summary} is devoted to summary and conclusion. 
In this paper, we assume a flat $\Lambda$CDM cosmological model with
the cosmological parameters : $\Omega_{\rm b}h^2=0.022$, $\Omega_{\rm m}h^2=0.13$,
$\Omega_\Lambda =0.72$, $h=0.7$, $n_{\rm s}=0.96$, $A_{\rm s}=2.4\times 10^{-9}$, and
$\tau =0.086$~\cite{Dunkley:2010ge}.
In Table \ref{notation}, we summarize the definition of the quantities
used to calculate the angular power spectrum.

\bc
\begin{table}[t]
\caption{
Notations for quantities used in this paper.
}
\vs{0.5}
\begin{tabular}{ccc} \hline \hline 
Symbol & eq. & Definition \\ 
\hline 
$\tilde g_{\mu\nu}$ & \eqref{eq:perturbed FLRW metric} 
& $4$-dimensional metric on conformal transformed spacetime\\
$g_{\mu\nu}$ & \eqref{eq:background metric} 
& $4$-dimensional metric on background spacetime\\
$\bar\gamma_{ij}$ & \eqref{eq:background metric} 
& $3$-dimensional spatial metric\\
$\omega_{ab}$, $\epsilon_{ab}$ & -
& Metric/Levi-Civita pseudo-tensor on unit sphere\\
\hline

semi-colon ( $;$ ) & - 
& Covariant derivative associated with $g_{\mu\nu}$\\
vertical bar ( $|$ ) & - 
& Covariant derivative associated with  $\bar\gamma_{ij}$\\
colon ( $:$ ) & \eqref{eq:intrinsic covariant derivative} 
& Covariant derivative associated with $\omega_{ab}$\\
$\nabla^2$ & - & Laplace operator on the unit sphere\\
\hline

$\hatn =(\theta ,\varphi )$ & - & Observed position on the sky \\
$k^\mu =E(1,-e^i_\chi)$ & \eqref{eq:k def} 
& Null vector on background spacetime \\
$E$, $e^i_\chi$ & - & Unperturbed photon energy and propagating direction \\
$u^\mu =(1,{\bm 0})$ & - 
& Observer's 4-velocity\\
$e^\mu_a =(0,e^i_a)$ & \eqref{eq:e^mu_a u^mu condition} 
& Orthonormal spacelike basis along light ray \\
$e^i_\pm$\,, $e^a_\pm =e_i^a\,e^i_\pm$ & \eqref{eq:polarization basis} 
& Basis of spin-weight $\pm 1$\\
$\lambda$, $\chi =E(\lambda_\rmO -\lambda )$ & -
& Affine parameter on background spacetime\\
\hline

$\sigma_{\rmg ,i}$ & \eqref{eq:sigma_rmg i def} 
& Gauge-invariant vector perturbations \\
$_0\sigma_\rmg =\sigma_{\rmg ,i}e^i_\chi$ & - 
& Spin-$0$ part of vector perturbations\\
$\sigma_{\rmg ,a}=\sigma_{\rmg ,i}e^i_a$ & - 
& Projected vector perturbations\\
$_{\pm 1}\sigma_\rmg =\sigma_{\rmg ,i}e^i_\pm$ & - 
& Spin-$\pm 1$ part of vector perturbations\\
$P_{\sigma_\rmg\sigma_\rmg}$ & \eqref{eq:power spectrum of vector perturbation def}
& Auto-power spectrum for $\sigma_\rmg$\\
\hline

$\xi^\mu$ & - & Deviation vector field\\
$\Delta^a$ & \eqref{eq:deflection angle} 
& Deflection angle on unit sphere \\
$x=\phi ,\varpi$ & \eqref{eq:gradient-curl decomposition} 
& Scalar/pseudo-scalar lensing potentials\\
$\mcD^a{}_b$ & \eqref{eq:Jacobi map def} 
& Jacobi map \\
$\mcT^a{}_b$ & \eqref{eq:Jacobi map eq} 
& Symmetric optical tidal matrix \\
$\gamma_{ab}$ & \eqref{eq:linear order Jacobi map solution}
& symmetric trace-free part of Jacobi map\\
$\gamma$, $g$ & \eqref{eq:shear def},\eqref{eq:reduced shear def} 
& Shear and reduced shear fields \\
$g^\rmX$ ($\rm X=\rmE ,\rmB$) & \eqref{eq:E-/B-mode def} 
& E-/B-mode reduced cosmic shear\\

$C_\ell^{\rm XX}$ & \eqref{eq:shear EE BB power spectrum},\eqref{eq:scalar and pseudo-scalar power spectrum} 
& Angular auto-power spectrum for X\\
$S^{\rm vector}_{{\rm X},\ell}$ &  \eqref{eq:E transfer function},\eqref{eq:B transfer function}
& Transfer function for E- and B-modes\\
$\epsilon_\ell^{(m)}$, $\beta_\ell^{(m)}$ & \eqref{eq:epsilon^0 def}-\eqref{eq:beta^1 def} 
& Radial E, B functions \\
$S^{\rm vector}_{x,\ell}$ &  \eqref{eq:phi transfer function},\eqref{eq:varpi transfer function} 
& Transfer function for scalar and pseudo-scalar lensing potential\\
\hline 
\end{tabular}
\label{notation}
\end{table} 
\ec

\section{Basic equations for weak lensing} \label{sec:Basic equations for weak lensing} 

In this section, we give the notation for the unperturbed and 
perturbed quantities, and derive the geodesic equation and geodesic deviation 
equations in the presence of vector metric perturbations. 
After the definitions of unperturbed and perturbed quantities in section 
\ref{sec:Pertubed universe},  
we give the basic equations which govern the gravitational 
lensing effect from vector perturbations and discuss the vector-induced
gravitational lensing effects in section \ref{sec:Geodesic equation} and
section \ref{sec:Geodesic deviation equation}.

\subsection{Perturbed universe}
\label{sec:Pertubed universe}

Throughout the paper we consider the flat 
Friedmann-Lema\^itre-Robertson-Walker (FLRW) universe 
with the metric given by 
\beq
	\dd s^2
		=a^2(\eta )\,\tilde g_{\mu\nu}\,\dd x^\mu\dd x^\nu
		=a^2(\eta )\,\Bigl( g_{\mu\nu}+h_{\mu\nu}\Bigr)\,\dd x^\mu\dd x^\nu
	\,,\label{eq:perturbed FLRW metric}
\eeq
where $a(\eta )$ corresponds to the conventional scale factor
of a homogeneous and isotropic universe, 
$h_{\mu\nu}$ is a small metric perturbation, $\tilde g_{\mu\nu}$
is the conformal flat metric which include the spacetime inhomogeneity,
\beq
	g_{\mu\nu}\dd x^\mu\dd x^\nu
		=-\dd\eta^2 +\bar\gamma_{ij}\dd x^i\dd x^j
		=-\dd\eta^2 +\dd\chi^2 +\chi^2\omega_{ab}\dd\theta^a\dd\theta^b
	\,,\label{eq:background metric}
\eeq
where $\omega_{ab}\dd\theta^a\dd\theta^b\equiv\dd\theta^2 +\sin^2\theta\dd\varphi$ 
is the metric on the unit sphere.

With the metric \eqref{eq:background metric},
vector perturbations are generally given by
\beq
	h_{00}=0
	\,,\ \ \ 
	h_{0i}=B_i
	\,,\ \ \ 
	h_{ij}=H_{i|j}+H_{j|i}
	\,,
\eeq
where both $B_i$ and $H_i$ are divergence-free three-vectors and 
the vertical bar ( $|$ ) denotes
the covariant derivative with respect to the three dimensional metric $\bar\gamma_{ij}$\,.
For convenience, we introduce the gauge-invariant
vector perturbations:
\beq
	\sigma_{\rmg ,i}\equiv\dot H_i-B_i
	\,.\label{eq:sigma_rmg i def}
\eeq
where the dot denotes the derivative with respect to the conformal time $\eta$.
Using the gauge freedom for the vector perturbations, 
we adopt the gauge $H_i=0$, so-called conformal Newton gauge, hereafter.
Appendix \ref{sec:Christoffel symbols and Riemann tensors from vector perturbations}
summarizes the Christoffel symbols and Riemann tensors from the vector
perturbations $h_{0i}=-\sigma_{\rmg ,i}$\,.
We now set the metric perturbations at the observer position to zero
because they can be absorbed into the homogeneous mapping.

We consider two geodesics $x^\mu (v)$ and 
$\tilde x^\mu (v)=x^\mu (v)+\xi^\mu (v)$\,, where $v$ is
the affine parameter and $\xi^\mu (v)$ is the deviation vector field.
Two geodesics lie in the past light cone of an observer $O$.
Since null geodesic is not affected by conformal transformations,
it is sufficient to consider the spacetime without the Hubble expansion.
Hence, we introduce a tangent vector $k^\mu$ along the geodesic $x^\mu (\lambda )$
on the conformally transformed spacetime with the affine parameter $\lambda$, 
defined by~\cite{Wald,Sasaki:1993tu}
\beq
	k^\mu\equiv a^2\frac{\dd x^\mu}{\dd v}
		=\frac{\dd x^\mu}{\dd\lambda}
	\,.\label{eq:k def}
\eeq
This null vector satisfies the equations:
\beq
	g_{\mu\nu}k^\mu k^\nu =0
	\,,\ \ \ 
	k^\mu{}_{;\nu}k^\nu
		=\frac{\dd^2 x^\mu}{\dd\lambda^2}
			+\Gamma^\mu_{\rho\sigma}
			\frac{\dd x^\rho}{\dd\lambda}\frac{\dd x^\sigma}{\dd\lambda}
		=0
	\,.
	\label{eq:conformally transformed geodesic eq}
\eeq
where the semi-colon ( $;$ ) and $\Gamma^\mu_{\rho\sigma}$ are
the covariant derivative and the Christoffel symbols associated with 
the unperturbed metric $g_{\mu\nu}$\,, 
respectively.
The geodesic equation at the zeroth-order in metric perturbations
reads that $x^\mu (\lambda )=E(\lambda ,(\lO -\lambda )\, e^i_\chi )$\,,
where $E$ and $e^i_\chi$ represent the photon energy and the photon
propagation direction measured from the observer in the background flat spacetime, 
$\lO$ denotes the affine parameter at $O$\,.
Note that the vector $e^i_\chi$ is the unit vector tangent to 
a geodesic on the flat three-space, satisfying 
$e_{\chi ,i} e^i_\chi =1$ and $(e^i_\chi )_{|j}e^j_\chi =0$. 
For convenience, we now switch from $\lambda$ to 
$\chi \equiv E(\lO -\lambda)$, hereafter.

Introducing the observer's 4-velocity at $O$, $u^\mu$,
we define orthonormal spacelike basis along the light ray, 
$e^\mu_a$ with $a=\theta,\varphi$, which satisfies
\beq
	g_{\mu\nu}e^\mu_a e^\nu_b =\omega_{ab}
	\,,\ \ \ 
	g_{\mu\nu}k^\mu e^\nu_a
		=g_{\mu\nu}u^\mu e^\nu_a =0
	\,.\label{eq:e^mu_a u^mu condition}
\eeq
They are parallely transported along the geodesics as
$u^\mu{}_{;\nu}k^\nu =0$, ${e^\mu_a}{}_{;\nu}k^\nu =0$\,.
For a static observer, 
we have $u^\mu =(1,{\bm 0})$ and $e^\mu_a =(0,e^i_a)$\,,
and the spatial basis vectors on the background spacetime 
in the Cartesian coordinate can be written as
\al{
	&e^i_\chi (\hatn )
		=\left(\sin\theta\cos\varphi ,\sin\theta\sin\varphi ,\cos\theta\right)
	\,,\\
	&e^i_\theta (\hatn )
		=\left(\cos\theta\cos\varphi ,\cos\theta\sin\varphi ,-\sin\theta\right)
	\,,\\
	&e^i_\varphi (\hatn )
		=\left( -\sin\theta\sin\varphi ,\sin\theta\cos\varphi ,0\right)
	\,,
}
where $\hatn =(\theta ,\varphi )$ is the observed position on the sky.
With these notations, we have
\al{
	e^i_\chi\,\pd_i =\pd_\chi
	\,,\ \ 
	e^i_\theta\,\pd_i =\frac{1}{\chi}\,\pd_\theta
	\,,\ \ 
	e^i_\varphi\,\pd_i =\frac{1}{\chi}\,\pd_\varphi
	\,,
}
and we can evaluate
\al{
	&\chi\, e^j_a\,\pd_j\, e^i_\chi 
		=e^i_a
	\,,\ \ 
	\chi^2\, e^j_{(a}\, e^k_{b)}\,\pd_j\,\pd_k\, e^i_\chi 
		=-\omega_{ab}\, e^i_\chi
	\,,\ \ 
	e^j_\chi\,\pd_j\, e^i_a =0
	\,,
	\notag\\
	&\chi\, e^j_\varphi\,\pd_j\, e^i_\theta
		=\chi\, e^j_\theta\,\pd_j\, e^i_\varphi
		=\cot\theta\, e^i_\varphi
	\,,\ 
	\chi\, e^j_\theta\,\pd_j\, e^i_\theta 
		=-e^i_\chi
	\,,\label{eq:basis derivative}\\
	&\chi\, e^j_\varphi\,\pd_j\, e^i_\varphi
		=-\sin\theta\,\left(\sin\theta\, e^i_\chi +\cos\theta\, e^i_\theta\right)
	\,.\notag
} 
We then define the intrinsic covariant derivative 
of a two-vector on the unit sphere, $X_a=X_ie^i_a$, in terms 
of the polarization basis as
\beq
	X_{a:b}\equiv \chi\, e^j_b\pd_j X_a -{}^{(2)}\Gamma^c_{ab}\,X_c
	\,,\ \ \ 
	^{(2)}\Gamma^c_{ab}\equiv \chi\, e^j_b\,e^c_i\,\pd_j\, e^i_a\, 
	\,.\label{eq:intrinsic covariant derivative}
\eeq
where $^{(2)}\Gamma^c_{ab}$ is the two dimensional Christoffel symbol defined 
on the unit sphere, 
and we have introduced the colon ( $:$ ) as the covariant derivative 
with respect to the unit sphere metric $\omega_{ab}$\,.
Here the polarization indices ($a,b,\cdots$) are raised or lowered
with respect to $\omega_{ab}$\,.

\subsection{Geodesic equation}
\label{sec:Geodesic equation}

At the linear-order in metric perturbations, the gravitational lensing effect appears 
on the spatial components of the geodesic equation for the photon ray.
The geodesic equation for the perturbed path 
$\tilde x^\mu (\chi )=x^\mu (\chi )+\xi^\mu (\chi )$
is given by 
\beq
	\frac{\dd^2\tilde x^\mu}{\dd\chi^2}
		+\tilde\Gamma^\mu_{\rho\sigma}\frac{\dd\tilde x^\rho}{\dd\chi}\frac{\dd\tilde x^\sigma}{\dd\chi}
	=0\,,\label{eq:conformally transformed geodesic eq2}
\eeq
where $\tilde\Gamma^\mu_{\rho\sigma}$ is the Christoffel symbols associated with 
the perturbed metric $\tilde g_{\mu\nu}$\,.
To derive the linear-order geodesic equation, we expand the Christoffel symbols as 
$\tilde\Gamma^\mu_{\rho\sigma}=\Gamma^\mu_{\rho\sigma}+\delta\Gamma^\mu_{\rho\sigma}$\,,
where $\Gamma^\mu_{\rho\sigma}$ is the Christoffel symbols associated with
the unperturbed metric $g_{\mu\nu}$
(see Appendix \ref{sec:Christoffel symbols and Riemann tensors from vector perturbations}).
The linear-order spatial geodesic equation for $\xi^i$ becomes
\beq
	\frac{\dd^2\xi^i}{\dd\chi^2}
		+2\Gamma^i_{jk}e^j_\chi\frac{\dd\xi^k}{\dd\chi}
		+\delta\Gamma^i_{\mu\nu}\frac{\dd x^\mu}{\dd\chi}\frac{\dd x^\nu}{\dd\chi}
		=0\,,\label{eq:perturbed spacelike geodesic eq}
\eeq
where $\dd /\dd\chi\equiv e^i_\chi \pd_i -\pd_\eta$\,.
To extract the angular components of the deviation vector, $\xi^a=\xi^ie_i^a$,
we multiply $e_i^a$ in both side of eq.~\eqref{eq:perturbed spacelike geodesic eq}\,.
Since the unperturbed Christoffel symbols 
satisfies $\Gamma^i_{jk}=0$ in the Cartesian coordinate system, with 
the condition for the parallel transportation, $(\dd /\dd\chi ) e_i^a=0$\,, 
we obtain the equation for $\xi^a$:
\beq
	\frac{\dd^2\xi^a}{\dd\chi^2}
		=\frac{1}{\chi}\omega^{ab}
			\biggl\{
				\left(_0\sigma_\rmg\right)_{:b}
				-\frac{\dd}{\dd\chi}\left(\chi\sigma_{\rmg ,b}\right)
			\biggr\}
	\,,\label{eq:eq_for_xi}
\eeq
where we have used eqs.~\eqref{eq:basis derivative}, \eqref{eq:perturbed Christoffel symbols},
and defined $_0\sigma_\rmg \equiv\sigma_{\rmg ,i}\, e^i_\chi$, 
$\sigma_{\rmg ,a}\equiv\sigma_{\rmg ,i}\, e^i_a$\,. 
Imposing the initial conditions, $\xi^a|_O=0$ and
$(\dd\xi^a /\dd\chi )|_O=\delta\theta^a_\rmO$, where $\delta\theta^a_\rmO$ denotes 
the angular coordinate at $O$, the solution for eq.~\eqref{eq:eq_for_xi} 
becomes
\al{
	\frac{\xi^a (\cS )}{\cS}
		=&\,\delta\theta^a_\rmO 
			+\omega^{ab}\int^{\cS}_0\frac{\dd\chi}{\chi}
				\biggl\{
					\frac{\cS -\chi}{\cS}\left(_0\sigma_\rmg\right)_{:b}
						-\sigma_{\rmg ,b}
				\biggr\}\biggl|_{(\eta_0 -\cS ,\cS e^i_\chi )}
	\,.\label{eq:linear order deflection angle solution}
}
In the above, the integral at the right-hand-side 
is evaluated along the unperturbed
light path according to the Born approximation.
We can confirm that this result exactly 
matches the one of Ref.~\cite{Schmidt:2012ne}.

Provided the deviation vector at the both end points, 
the deflection angle, $\Delta^a$\,, can be estimated 
through~\cite{Stebbins:1996wx}
\beq
	\Delta^a
		\equiv\frac{\xi^a (\cS )}{\cS} -\delta\theta^a_\rmO
	\,.\label{eq:deflection angle}
\eeq
The deflection angle $\Delta^a$ is the two-dimensional 
vector field defined on a celestial sphere, and it can be uniquely 
characterized by introducing two potentials; scalar ($\phi$) and 
pseudo-scalar ($\varpi$) lensing potentials.  
Then, the deflection angle is described by the sum of 
the two terms (e.g., \cite{Hirata:2003ka}):
\beq
	\Delta_a =\phi_{:a}+\varpi_{:b}\,\epsilon^b{}_a
	\,,\label{eq:gradient-curl decomposition}
\eeq
where $\epsilon^b{}_a$ denotes the two dimensional Levi-Civita pseudo-tensor.
Hereafter, we call the first and second terms in the right-hand-side of 
eq.~(\ref{eq:gradient-curl decomposition}) 
gradient- and curl-modes, respectively ~\cite{Namikawa:2011cs}. 
The scalar-/pseudo-scalar lensing potentials 
in the case of the vector perturbations can be written as
\al{
	&\nabla^2\phi
		=\Delta^a{}_{:a}
		=\int^{\cS}_0\frac{\dd\chi}{\chi}
				\biggl\{
					\frac{\cS -\chi}{\cS}\nabla^2\left(_0\sigma_\rmg\right)
					-\sigma_\rmg{}^a{}_{:a}
				\biggr\}
	\,,\label{eq:gradient mode induced by vector}\\
	&\nabla^2\varpi
		=\Delta^a{}_{:b}\,\epsilon^b{}_a
		=-\int^{\cS}_0\frac{\dd\chi}{\chi}\sigma_\rmg{}^a{}_{:b}\,\epsilon^b{}_a
	\,,\label{eq:curl mode induced by vector}
}
where $\nabla^2$ is the Laplace operator on the sphere, namely 
$\nabla^2\phi =\phi_{:ab}\,\omega^{ab}=\phi\,^{:a}{}_{:a}$\,. 
Note that eq.~\eqref{eq:curl mode induced by vector} coincides
with eq.~(1.4) of Ref.~\cite{Namikawa:2011cs}\,.
The curl component of the deflection becomes non-vanishing in the presence
of the divergence-free component of $\sigma_{\rmg ,a}$\,.
In section \ref{sec:Deflection angle}, using 
eqs.~\eqref{eq:gradient mode induced by vector} and 
\eqref{eq:curl mode induced by vector}, 
we will derive the explicit formulas for the gradient-/curl-mode deflection angle
and their angular power spectrum.

\subsection{Geodesic deviation equation}
\label{sec:Geodesic deviation equation}

Here, we introduce the Jacobi map which characterizes
the deformation of light bundle.
In terms of the projected deviation vector $\xi^a$\,,
the geodesic deviation equation in the conformal transformed spacetime 
can be written as~\cite{Seitz:1994xf,1961RSPSA.264..309S}
\beq
	\frac{\dd^2\xi^a}{\dd\chi^2}=\mcT^a{}_b\,\xi^b\, ;\ 
	\mcT^a{}_b=-\frac{1}{E^2}R_{\mu\rho\nu\sigma}k^\mu k^\nu e^{\rho a} e^\sigma_b
	\,,
	\label{eq:deviation eq}
\eeq
where $\mcT^a{}_b$ is called the symmetric optical tidal matrix,
and $R_{\mu\rho\sigma\nu}$ is the Riemann tensor of the metric $g_{\mu\nu}$\,.
Given the initial conditions at the observer, 
$\xi^a|_O=0$ and $(\dd\xi^a /\dd\chi )|_O=\delta\theta^a_\rmO$\,, 
the solution of eq.~\eqref{eq:deviation eq} is generally rewritten in
the following form:
\beq
	\xi^a (\chi )\equiv\mcD^a{}_b(\chi )\,\delta\theta^b_\rmO
	\,,
	\label{eq:Jacobi map def}
\eeq
where $\mcD^a{}_b$ is the Jacobi map and it satisfies
\beq
	\frac{\dd^2}{\dd\chi^2}\mcD^a{}_b=\mcT^a{}_c\mcD^c{}_b
	\,,
	\label{eq:Jacobi map eq}
\eeq
with the initial condition at the observer $O$\, rewritten with  
$\mcD^a{}_b|_O=0$ and $(\dd /\dd\chi )\mcD^a{}_b|_O=\delta^a{}_b$\,.

We are particularly concerned with the perturbed Jacobi matrix induced by 
vector perturbations. To get the expressions relevant for the weak lensing
measurements, we expand eq.~\eqref{eq:Jacobi map eq} 
as $\mcD^a{}_b=\bar\mcD^a{}_b+\delta\mcD^a{}_b$\,,
and $\mcT^a{}_b=\bar\mcT^a{}_b+\delta\mcT^a{}_b$\,.
Since the tidal matrix vanishes 
in the unperturbed spacetime, $\bar\mcT^a{}_b=0$, 
the zeroth-order solution of Jacobi map becomes 
$\bar\mcD^a{}_b =\chi\delta^a{}_b$\,. 
Plugging this expression into eq.~\eqref{eq:Jacobi map eq} and 
solving the linear-order equation, we 
obain the expression valid up to the linear-order in metric 
perturbations:  
\beq
	\mcD^a{}_b(\cS )
		=\cS\,\delta^a{}_b
			+\int^{\cS}_0\dd\chi\,\left(\cS -\chi\right)\chi\,
                        \delta\mcT^a{}_b(\chi )
			+\mcO (h^2)
	\,.\label{eq:Jacobi map perturbed formal solution}
\eeq
In the above, 
an important observation is that the resultant Jacobi map is always 
symmetric. Hence, the anti-symmetric part of the Jacobi map, 
which is directly related to the rotation mode, does not appear at the 
linear order.

Note that Eq.~(\ref{eq:Jacobi map perturbed formal solution}) is 
still general in the sense that the Jacobi map 
includes the deformation arising from all of the metric 
perturbations. To specifically derive 
the expression relevant for the vector perturbations, we 
explicitly write down the tidal matrix. 
Using eqs.~\eqref{eq:linearized optical tidal matrix}, 
\eqref{eq:basis derivative}, and 
\eqref{eq:intrinsic covariant derivative},
we have
\al{
	\delta\mcT_{ab}\Bigl|_{\rm vector}
		=&\, -e^i_\chi e^j_{(a}e^k_{b)}\pd_j\pd_k h_{0i} 
				+\frac{\dd}{\dd\chi}\left( e^i_{(a}e^j_{b)}\pd_j h_{0i}\right)
	\notag\\
		=&\,\frac{1}{\chi^2}
			\left\{
				\left(_0\sigma_\rmg\right)_{:ab}
				-\frac{\dd}{\dd\chi}\left(\chi\,\sigma_{\rmg ,(a:b)}\right)
				+\chi\,\omega_{ab}\left({}_0\dot\sigma_\rmg\right)
			\right\}
	\,,
}
where we have introduced the symmetric operation defined 
by $A_{(a}B_{b)}\equiv \frac{1}{2}(A_aB_b+A_bB_a)$\,.

Since we are interested in the shear fields, it is sufficient to consider
the symmetric trace-free part of the Jacobi map, $\mcD_{\ave{ab}}$, where
the angle bracket $\ave{\cdots}$ denotes the symmetric trace-free part taken
in the two-dimensional space: 
$X_{\ave{ab}}\equiv (X_{ab}+X_{ba}-X^c{}_c\,\omega_{ab})/2$\,.
For the vector perturbations, the symmetric trace-free part of 
eq.~\eqref{eq:Jacobi map perturbed formal solution} reduces to
(see also \cite{Bernardeau:2011tc,Bernardeau:2009bm,Schmidt:2012ne})
\al{
	\gamma_{ab}\equiv\frac{1}{\cS}\mcD_{\ave{ab}}(\cS )
		=&\,\int^{\cS}_0\frac{\dd\chi}{\chi}
				\biggl\{
					\frac{\cS -\chi}{\cS}\left(_0\sigma_\rmg\right)_{:\ave{ab}}
					-\sigma_{\rmg ,\ave{a:b}}
				\biggr\}\biggl|_{(\eta_0 -\cS ,\cS e^i_\chi )}
	\,.\label{eq:linear order Jacobi map solution}
}
Strictly speaking, the affine parameter $\lambda$ and/or 
the conformal distance $\chi$ are not direct observables, and 
we should express the Jacobi map as a function of the redshift 
of the source, $z_\rmS$\,, taking 
the perturbation of the observed redshift into account. 
At first-order, however, 
the perturbation of the redshift affects only the trace part 
of the Jacobi map~\cite{Bernardeau:2009bm}. 
Thus, simply relating $\chi$ with redshift $z$ through 
$\chi=\int^z_0 dz'/H(z')$, 
the expression \eqref{eq:linear order Jacobi map solution} 
still remains relevant, and we will use it to derive the formulas for 
angular power spectra of the E-/B-mode cosmic shear.

Finally, we note that 
eqs.~\eqref{eq:linear order deflection angle solution} or 
\eqref{eq:deflection angle} and \eqref{eq:linear order Jacobi map solution} 
lead to the following simple relation: 
\beq
	\gamma_{ab}=\Delta_{\ave{a:b}}
	\,.\label{eq:Jacobi map-deflection angle relation}
\eeq
This relation is valid at linear order, and exactly coincides with the 
one empirically defined in \cite{Stebbins:1996wx}. On the other hand, 
no such expression is obtained for the relation between 
$\mcD^a{}_b$ and $\Delta^a{}_{:b}$ because of the non-vanishing trace part.

\section{Weak lensing observables induced by vector perturbations} 
\label{sec:Weak lensing observations induced by vector perturbations} 

In this section, we derive the full-sky formulas for
the angular power spectra of the E-/B-mode cosmic shear
and the gradient-/curl-mode deflection angle generated
by vector metric perturbations. The formula for the angular 
power spectra are respectively given in 
section \ref{sec:Cosmic shear} and \ref{sec:Deflection angle}, 
for the E-/B-mode cosmic shear 
[eqs.~\eqref{eq:shear EE BB power spectrum}, 
\eqref{eq:E transfer function}--\eqref{eq:beta^1 def}], 
and the gradient-/curl-mode deflection angle
[eqs.~\eqref{eq:scalar and pseudo-scalar power spectrum},
\eqref{eq:phi transfer function}, and \eqref{eq:varpi transfer function}].  
We then discuss an interesting relation for angular power spectra
between the deflection angle 
and the cosmic shear in section \ref{sec:Shear-deflection relation}.  

\subsection{Cosmic shear} \label{sec:Cosmic shear}

In discussing the spatial patterns of shear fields on celestial sphere, 
it is useful to introduce the spin-weighted quantities. 
A quantity $_sX$ that transforms as $_sX\rightarrow e^{i\alpha s}\,_sX$ under 
a rotation of $(e^i_\theta\,, e^i_\varphi )$ by an angle $\alpha$ 
is called spin-weighted quantity with spin-$s$\,.
According to \cite{Lewis:2001hp}, we define the 
basis of spin-weight $\pm 1$ as
\beq 
	e^i_\pm (\hatn )
		\equiv e^i_\theta (\hatn )\pm\frac{i}{\sin\theta}e^i_\varphi (\hatn )
	\,.\label{eq:polarization basis}
\eeq 
With this basis, the Jacobi map can be decomposed 
into the spin-$0$ and spin-$\pm 2$ components~\cite{Bernardeau:2009bm}:
\beq 
	_0\mcD =\mcD_{ab}e^a_+e^b_-
	\,,\ \ \ 
	_{\pm 2}\mcD =\mcD_{ab}e^a_\pm e^b_\pm
	\,,
\eeq 
where we have defined the projected basis $e^a_\pm\equiv e^a_ie^i_\pm$\,. 
As we mentioned in section \ref{sec:Geodesic deviation equation}, 
the Jacobi map at the linear order is symmetric, 
and the spin-$0$ part of the linear-order Jacobi map contains only the 
trace part, which is related to the convergence field $\kappa$ :
\beq
	_0\mcD 
		=\Tr\,\mcD^a{}_b 
		\equiv 2\cS\left( 1-\kappa\right)
	\,.
\eeq
On the other hand, 
the spin-$\pm 2$ parts give the shear fields, $\gamma$ and $\gamma^*$,
defined by
\beq
	\gamma  
		\equiv -\frac{_{+2}\mcD}{2\cS}
		=-\frac{1}{2}\gamma_{ab}e^a_+e^b_+
	\,,\ \ \ 
	\gamma^*
		\equiv -\frac{_{-2}\mcD}{2\cS}
		=-\frac{1}{2}\gamma_{ab}e^a_-e^b_-
	\,.\label{eq:shear def}
\eeq
In practice, what we observe is not directly the shear itself, but rather 
the ratio between the anisotropic and isotropic deformations,
so-called reduced shear.
The reduced shear is related to the spin-weighted Jacobi map as
\beq 
	g\equiv\frac{\gamma}{1-\kappa}
		=-\frac{_{+2}\mcD}{_0\mcD}
	\,,\ \ \ 
	g^*\equiv -\frac{_{-2}\mcD}{_0\mcD}
	\,.\label{eq:reduced shear def}
\eeq 
Note that at linear-order, the reduced shear fields is simply
described by the shear fields, i.e., $g\simeq\gamma$ and $g^*\simeq\gamma^*$.
Since the reduced shear fields transform as the spin-$\pm 2$ 
quantities, they are decomposed on the basis of spin-$\pm 2$ 
harmonics ${}_{\pm 2}Y_{\ell m}(\hatn )$ 
(see appendix \ref{sec:Spin operators and spin-weighted spherical harmonics} 
for definition):
\beq
	g(\hatn )
		=\sum_{\ell =2}^\infty\sum_{m=-\ell}^{+\ell}\,{}_{+2}
				g_{\ell m}\,{}_{+2}Y_{\ell m}(\hatn )
	\,,\ \ \  
	g^*(\hatn )
		=\sum_{\ell =2}^\infty\sum_{m=-\ell}^{+\ell}\,{}_{-2}
				g_{\ell m}\,\,{}_{-2}Y_{\ell m}(\hatn )
	\,.\label{eq:spherical expansion for reduced shear}
\eeq
Then, E- and B-modes, as the two parity eigenstates, can be defined: 
\beq 
	g_{\ell m}^\rmE 
		=-\frac{1}{2}\Bigl({}_{+2}g_{\ell m}+{}_{-2}g_{\ell m}\Bigr)
	\,,\ \ \ 
	g_{\ell m}^\rmB 
		=-\frac{1}{2i}\Bigl({}_{+2}g_{\ell m}-{}_{-2}g_{\ell m}\Bigr)
	\,.\label{eq:E-/B-mode def}
\eeq 
The auto- and cross-power spectra of these quantities are also defined as:
\beq 
	C_\ell^{\rm XX'}\equiv\frac{1}{2\ell +1}\sum_m\ave{g^{\rm X*}_{\ell m}\,g^{\rm X'}_{\ell m}}
	\,,\label{eq:def_C_ell_EB}
\eeq 
where ${\rm X},{\rm X}'=E,B$, and the angle bracket $\ave{\cdots}$ denotes 
the ensemble average.

With the preliminary setup mentioned above, 
let us now derive the explicit expression 
for E-/B-mode power spectra. Multiplying $e^a_+e^b_+$ in both side of 
eq.~\eqref{eq:linear order Jacobi map solution}, 
the spin-$+2$ part of the reduced shear, eq.~\eqref{eq:reduced shear def}, 
is written as 
(see also \cite{Bernardeau:2011tc,Bernardeau:2009bm})
\al{
	g=&-\frac{1}{2}\int^{\cS}_0\frac{\dd\chi}{\chi}
				\biggl\{
					\frac{\cS -\chi}{\cS}\left(_0\sigma_\rmg\right)_{:ab}e^a_+e^b_+
						-\left(\sigma_{\rmg ,a:b}\right) e^a_+e^b_+
				\biggr\}
	\,.\label{eq:reduced shear from vector perturbations}
}
In the above, the gauge-invariant vector perturbation 
contains statistical information for 
spatial randomness, which can be decomposed into the Fourier modes.  
For a given Fourier mode ${\bm k}$\,, a convenient representation 
would be $e^i_\pm (\hat{\bm k})\,e^{-i{\bm k}\cdot{\bm r}}$,
where $\hat{\bm k}={\bm k}/k$, and  
$e^i_\pm (\hat{\bm k})$ is the basis vector perpendicular to $\hat{\bm k}$\,.
We then write $\sigma_\rmg{}^i$ as 
\beq 
	\sigma_\rmg{}^i({\bm r} ,\eta )
		=\3Dint{\bm k} e^{-i{\bm k}\cdot{\bm r}}
			\biggl\{
				\sigma_\rmg^{(+1)}({\bm k},\eta )e_+^i(\hatk )
				+\sigma_\rmg^{(-1)}({\bm k},\eta )e_-^i(\hatk )
			\biggr\}
    \,.\label{eq:Fourier exp sigma}
\eeq 
We note that the trace-free condition for the vector perturbations, 
namely $\sigma_\rmg{}^i{}_{|i}=0$\,, is automatically satisfied
when we consider the Fourier expansion, eq.~\eqref{eq:Fourier exp sigma}.
The quantities, $\sigma_\rmg^{(+1)}({\bm k},\eta )$ and $\sigma_\rmg^{(-1)}({\bm k},\eta)$, 
are responsible for the randomness 
arising from initial condition and/or late-time
evolution. Assuming the un-polarized state of vector fluctuations, 
their statistical properties are characterized by
\al{
	&\ave{
		{\sigma^{(s)}_\rmg}^*({\bm k},\eta )\sigma^{(s')}_\rmg ({\bm k}',\eta' )
		}
		=\Biggl\{
			\begin{array}{ll}
				\frac{1}{2}P_{\sigma_\rmg\sigma_\rmg}(k;\eta ,\eta' )\,(2\pi )^3\delta^{3}({\bm k}-{\bm k}') & :\  s=s'=\pm 1\\
				0 & :\  s\neq s'\\
			\end{array}
	\,.\label{eq:power spectrum of vector perturbation def}
}
Substituting eq.\eqref{eq:Fourier exp sigma} into the 
expression \eqref{eq:reduced shear from vector perturbations}, we 
first explore the relation between the 
harmonic coefficients $g_{\ell m}^{\rm E,B}$ and
$\sigma^{(\pm1)}_\rmg$. Next plugging this into the definition 
\eqref{eq:def_C_ell_EB}, after lengthy calculation presented in appendix 
\ref{sec:E-/B-modes of shear fields}, we obtain
the formulas for the power spectrum of E-/B-mode cosmic shear. The 
resultant expressions become   
\al{ 
	&C_\ell^{\rmX\rmX}
		=\frac{2}{\pi}
			\int^\infty_0 k^2\dd k\int^{\cS}_0 k\,\dd\chi\int^{\cS}_0 k\,\dd\chi'
			S^{\rm vector}_{\rmX ,\ell}(k,\chi )S^{\rm vector}_{\rmX ,\ell}(k,\chi' )
			P_{\sigma_\rmg\sigma_\rmg}\left( k;\eta_0 -\chi ,\eta_0 -\chi'\right)
	\,,\label{eq:shear EE BB power spectrum}\\
	&C_\ell^{\rm EB}=0
	\,,
}
where the quantities $S^{\rm vector}_{\rmX ,\ell}$ are the transfer functions 
defined as
\al{
	&S^{\rm vector}_{\rmE ,\ell}(k,\chi )
		=\frac{\cS -\chi}{\cS}\epsilon^{(0)}_\ell (k\chi )-\epsilon^{(1)}_\ell (k\chi )
	\,,\label{eq:E transfer function}\\
	&S^{\rm vector}_{\rmB ,\ell}(k,\chi )
		=\beta^{(1)}_\ell (k\chi )
	\,\label{eq:B transfer function}
}
with the coefficients $\epsilon^{(0,1)}_\ell\,,\beta^{(1)}_\ell$ given by
(see also \cite{Hu:1997hp})
\al{
	\epsilon^{(0)}_\ell (x)
		=&\frac{1}{2}\sqrt{\frac{(\ell -1)!}{(\ell +1)!}\frac{(\ell +2)!}{(\ell -2)!}}
			\,\ell (\ell +1)\frac{j_\ell (x)}{x^2}
	\,,\label{eq:epsilon^0 def}\\
	\epsilon^{(1)}_\ell (x)
		=&\frac{1}{2}\sqrt{\frac{(\ell -1)!}{(\ell +1)!}\frac{(\ell +2)!}{(\ell -2)!}}
			\,\left(\frac{j_\ell (x)}{x^2}+\frac{j'_\ell (x)}{x}\right)
	\,,\label{eq:epsilon^1 def}\\
	\beta^{(1)}_\ell (x)
		=&\frac{1}{2}\sqrt{\frac{(\ell -1)!}{(\ell +1)!}\frac{(\ell +2)!}{(\ell -2)!}}\,
			\frac{j_\ell (x)}{x}
	\,.\label{eq:beta^1 def}
}
Here, $j_\ell (x)$ is the spherical Bessel function.
These formulas 
[eqs.\eqref{eq:shear EE BB power spectrum}-\eqref{eq:beta^1 def}] 
are one of the main results of this paper. 

\subsection{Deflection angle} \label{sec:Deflection angle}

Based on the expressions~\eqref{eq:gradient mode induced by vector} 
and \eqref{eq:curl mode induced by vector}, 
let us next consider the deflection angle.  
First notice that the metric on the sphere, $\omega^{ab}$, and the Levi-Civita
pseudo-tensor, $\epsilon^{ab}$, 
can be rewritten in term of the basis vectors $e^a_\pm$ 
with~\cite{Lewis:2006fu}
\beq
	\omega^{ab}=e^{(a}_+e^{b)}_-
	\,,\ \ \ 
	\epsilon^{ab}=i\, e^{[a}_+e^{b]}_-
	\,,\label{eq:omega-epsilon-spin basis relations}
\eeq
where we have introduced the anti-symmetric operation defined by 
$A_{[a}B_{b]}=\frac{1}{2}(A_{a}B_{b}-A_{b}B_a)$\,.
Then, the scalar and pseudo-scalar lensing potentials are recast as
\al{
	\nabla^2\phi
		=&\int^{\cS}_0\frac{\dd\chi}{\chi}
			\biggl\{
				\frac{\cS -\chi}{\cS}\left(_0\sigma_\rmg\right)_{:ab}e^a_+e^b_-
				-\,\sigma_{\rmg ,a:b}e^{(a}_+e^{b)}_-
			\biggr\}
	\,,\label{eq:scalar lensing potential in spin op}\\
	\nabla^2\varpi 
		=&-\,i\int^{\cS}_0\frac{\dd\chi}{\chi}\sigma_{\rmg ,a:b}e^{[a}_-e^{b]}_+
	\,.\label{eq:pseudo-scalar lensing potential in spin op}
}
Since the scalar/pseudo-scalar lensing potentials, $\phi$ and $\varpi$, transform 
as spin-$0$ quantities, they are decomposed with the spin-$0$ harmonics: 
\beq
	\phi (\hatn )=\sum_{\ell =1}^\infty\sum_{m=-\ell}^\ell\phi_{\ell m}\,Y_{\ell m}(\hatn )
	\,,\ \ \ 
	\varpi (\hatn )=\sum_{\ell =1}^\infty\sum_{m=-\ell}^\ell\varpi_{\ell m}\,Y_{\ell m}(\hatn )
	\,.\label{eq:phi varpi decomposition}
\eeq
The angular power spectra for the scalar/pseudo-scalar lensing potentials 
are defined as
\beq
	C_\ell^{xx'}=\frac{1}{2\ell +1}\sum_m\ave{x_{\ell m}^*x'_{\ell m}}
	\,,\label{eq:def_C_ell_phi}
\eeq
with $x,x'=\phi ,\varpi$\,.

Now, similar manner to the cosmic shear, we first derive the relation 
between $x_{\ell m}$ and $\sigma_\rmg^{(\pm1)}$. This is done with 
a rather lengthy calculation in 
appendix \ref{sec:Scalar/pseudo-scalar lensing potential}. 
The resultant expressions are summarized in equations 
(\ref{eq:phi_ell_m_sigma}) and (\ref{eq:varpi_ell_m_sigma}). 
Then, the angular power spectra for the gradient- and curl-modes
are obtained by plugging these expressions into \eqref{eq:def_C_ell_phi}. 
As a result, we have 
\al{
	C_\ell^{xx}
		=&\frac{2}{\pi}\int^\infty_0 k^2\dd k
			\int^{\cS}_0\, k\,\dd\chi\int^{\cS}_0\, k\,\dd\chi'
			S^{\rm vector}_{x,\ell}(k,\chi )S^{\rm vector}_{x,\ell}(k,\chi' )
			P_{\sigma_\rmg\sigma_\rmg}\left( k;\eta_0 -\chi ,\eta_0 -\chi'\right)
	\,,\label{eq:scalar and pseudo-scalar power spectrum}\\
	C_\ell^{\phi\varpi}=&0
	\,,
}
where $P_{\sigma_\rmg\sigma_\rmg}$ was defined in 
eq.~\eqref{eq:power spectrum of vector perturbation def}. 
The transfer functions $S^{\rm vector}_{x,\ell}$ are defined by
\al{
	&S^{\rm vector}_{\phi ,\ell}(k,\chi )
		=2\,\sqrt{\frac{(\ell -2)!}{(\ell +2)!}}\,
		 \left(
			\frac{\cS -\chi}{\cS}\,\epsilon^{(0)}_\ell (k\chi )
			-\epsilon^{(1)}_\ell (k\chi )
		 \right)
	\,,\label{eq:phi transfer function}\\
	&S^{\rm vector}_{\varpi ,\ell}(k,\chi )
		=2\,\sqrt{\frac{(\ell -2)!}{(\ell +2)!}}\,\beta^{(1)}_\ell (k\chi )
	\,,\label{eq:varpi transfer function}
}
with the quantities $\epsilon_\ell^{(0,1)}$ and $\beta_\ell^{(1)}$ 
given by 
eqs.~\eqref{eq:epsilon^0 def}-\eqref{eq:beta^1 def}. 
The formulas given above 
are also one of the main results of this paper.

\subsection{Shear-deflection relation} \label{sec:Shear-deflection relation}

Here, we briefly mention the relation between E-/B-mode cosmic shear and 
gradient-/curl-mode deflection angle. From 
eq.~\eqref{eq:Jacobi map-deflection angle relation},  
we found that in the case of vector perturbations 
the symmetric trace-free part of the Jacobi map, namely the shear fields,
are directly related to the deflection angle, which can be decomposed into
the gradient and curl modes
[see eq.~\eqref{eq:gradient-curl decomposition}]:
\beq 
	\Delta_{\ave{a:b}}
		=\phi_{:\ave{ab}}+\varpi_{:c\langle a}\,\epsilon^c{}_{b\rangle}
		=\gamma_{ab}
	\,.\label{eq:Delta-gamma relation}
\eeq 
This implies that we can extract the vector-induced 
gradient-/curl-modes from the vector-induced shear fields.
Taking the divergence and then taking the divergence or curl again on 
eq.~\eqref{eq:Delta-gamma relation}, 
we find~\cite{Stebbins:1996wx,Lewis:2006fu}
\beq 
	\gamma_{ab}{}^{:ab}
		=\frac{1}{2}\nabla^2\left(\nabla^2 +2\right)\phi
	\,,\ \ \ 
	\epsilon^{ca}\gamma_{ab}{}^{:b}{}_{:c}
		=\frac{1}{2}\nabla^2\left(\nabla^2 +2\right)\varpi
  \,.\label{eq:shear-lensing potential relations}
\eeq 
The above relation
can be further reduced to simplified forms if we move to the harmonic space. 
After lengthy calculation presented in appendix 
\ref{sec:Derivation of shear-deflection relation}\,,
we obtain the explicit relation between $\phi_{\ell m}$\,, $\varpi_{\ell m}$ 
defined in eq.~\eqref{eq:phi varpi decomposition} and
$g_{\ell m}^\rmE$\,, $g_{\ell m}^\rmB$ 
defined in eq.~\eqref{eq:E-/B-mode def} as
\al{
	\phi_{\ell m}
		=2\sqrt{\frac{(\ell -2)!}{(\ell +2)!}}\,g_{\ell m}^\rmE
	\,,\ \ \ 
	\varpi_{\ell m}
		=2\sqrt{\frac{(\ell -2)!}{(\ell +2)!}}\,g_{\ell m}^\rmB
	\,.\label{eq:phi varpi - g^E g^B}
} 
Thus, we reach at the relation between angular power spectra for 
shear and deflection angle: 
\al{
	&C^{\phi\phi}_\ell =4\frac{(\ell -2)!}{(\ell +2)!}C_\ell^{\rm EE}
	\,,\ \ \ 
	C^{\varpi\varpi}_\ell =4\frac{(\ell -2)!}{(\ell +2)!}C_\ell^{\rm BB}
	\,,\ \ \ 
	C^{\phi\varpi}_\ell =C_\ell^{\rm EB}=0
	\,.\label{eq:relation between phi/varpi-EE/BB}
}
These relations are nontrivial, 
but are generally valid as long as the relation between the deflection angle
and the Jacobi map, eq.~\eqref{eq:Delta-gamma relation}, holds.
Note that the relations given here does not 
hold for general distribution of background sources, 
if the source distribution for shear fields 
differs from that for the deflection angle. 

\section{Implications for cosmic string network} 
\label{sec:Implications to a cosmic string network} 

In this section, as an illustrative example 
for the application of the full-sky formulas, 
we consider a cosmic string network 
as a possible source for seeding vector perturbations. 
Based on a simple model of cosmic strings described in Sec.
\ref{sec:Vector perturbations generated by a cosmic string network}, 
signal-to-noise ratios for B-mode cosmic shear and curl-mode 
deflection angle are estimated, and the detectability of the string 
network from future observations 
is discussed in Sec.~\ref{sec:Observational implications}. 

\subsection{Vector perturbations generated by a cosmic string network} 
\label{sec:Vector perturbations generated by a cosmic string network}

In most of the active generation mechanisms, 
the vector metric perturbations are sourced by  
the non-vanishing vector mode of stress-energy tensor. 
This is true for the case of cosmic string network. 
Thus, to compute the weak lensing signals,  
we must first evaluate the vector stress-energy 
tensor induced by the cosmic strings. 
Let us write the vector stress-energy as \cite{Hu:1997hp}:
\beq
	\delta T^0{}_i({\bm r},\eta )
		=\int\frac{\dd^{3}{\bm k}}{(2\pi )^3}e^{-i{\bm k}\cdot{\bm r}}
			\biggl\{v^{(+1)}({\bm k},\eta )e^i_+(\hat{\bm k})
				+v^{(-1)}({\bm k},\eta )e^i_-(\hat{\bm k})\biggr\}
\,.\label{eq:velocity perturbations}
\eeq
Through the linearized Einstein equation, this is related to the vector 
metric perturbation as~\cite{Hu:1997hp}
\beq
	\sigma_\rmg^{(\pm 1)}({\bm k},\eta )
		=\frac{16\pi Ga^2}{k^2}v^{(\pm 1)}({\bm k},\eta )
	\,.\label{eq:linearized Einstein eq}
\eeq
Here, we have ignored the contribution of the anisotropic stress tensor 
from the cosmological fluids.

For a concrete model of non-vanishing stress-energy tensor, 
we consider a string network described by the velocity-dependent one-scale (VOS) 
model~\cite{Martins:2000cs,Martins:1996jp,Avgoustidis:2005nv,Takahashi:2008ui}.
The string network consists of a collection of string segments, 
whose length and velocity are 
respectively given by 
$\xi=1/(H\gamma_{\rm s})$ and $v_{\rm rms}$, where $\gamma_{\rm s}$ 
represents the correlation length of the string network. 
The string segments are assumed to be randomly oriented and Poisson-distributed.
The intercommuting process provides an essential mechanism for
a string network to lose its energy due to loop formation. 
It is widely believed that the energy-loss mechanism allows
the network to relax towards an cosmological 
attractor solution, in which $\gamma_{\rm s}$ and $v_{\rm rms}$ remain constant:
this is so-called scaling solution.
Several groups developed the numerical codes to evolve a string network
and concluded that there exists a scaling regime for long strings~\cite{Bevis:2010gj,Bevis:2007qz,Albrecht:1984xv,Albrecht:1989mk,Bennett:1987vf,Bennett:1989yp,Ringeval:2005kr}.
In the VOS model, $\gamma_{\rm s}$ and $v_{\rm rms}$ are approximately 
described by $\gamma_{\rm s} =(\pi\sqrt{2}/3\tilde{c}\,P)^{1/2}$ 
and $v_{\rm rms}^2=(1-\pi /3\gamma_{\rm s} )/2$~\cite{Takahashi:2008ui}, 
where $\tilde c\approx 0.23$ quantifies the efficiency of loop formation~\cite{Martins:2000cs} 
and $P$ is the intercommuting probability.
For a tractable analytic estimate, we assume that the correlations 
between the string segments are characterized by the simple model developed 
in \cite{Vincent:1996qr,Albrecht:1997mz,Hindmarsh:1993pu}. Then, from 
Appendix \ref{sec:Derivation of correlations of a cosmic string network}, 
we obtain the equal-time auto power spectrum for the vector perturbations, 
$P_{\sigma_\rmg\sigma_\rmg}(k;\eta,\eta)$:  
\beq
	P_{\sigma_\rmg\sigma_\rmg}(k;\eta ,\eta )
		=\left( 16\pi G\mu\right)^2
					\frac{2\sqrt{6\pi}\, v_{\rm rms}^2}{3(1-v_{\rm rms}^2)}
					\frac{4\pi\chi^2 a^4}{H}\left(\frac{a}{k\xi}\right)^5
					{\rm erf}\left(\frac{k\xi /a}{2\sqrt{6}}\right)
	\,,\label{eq:equal-time auto-power spectrum}
\eeq
where ${\rm erf}(x)$ is the error function, 
${\rm erf}(x)=(2/\sqrt{\pi})\int^x_0\dd y\, e^{-y^2}$. 
To compute the weak lensing power spectra, 
we further need the unequal-time auto-power spectrum. Here, 
as a crude estimate, we adopt the 
following approximation \cite{Magueijo:1995xj,Durrer:1998rw,Durrer:2001xu}: 
\beq
	P_{\sigma_\rmg\sigma_\rmg}(k;\eta_1 ,\eta_2 )
		=\,\sqrt{
				P_{\sigma_\rmg\sigma_\rmg}(k,\eta_1 ,\eta_1 )
				P_{\sigma_\rmg\sigma_\rmg}(k,\eta_2 ,\eta_2 )
			}
		\,.\label{eq:perfectly coherent condition}
\eeq

The model and assumptions given above would be simplistic and 
might not be realistic for a precision study of lensing signals. 
Further, eqs.~(\ref{eq:equal-time auto-power spectrum}) and 
(\ref{eq:perfectly coherent condition}) may have additional modifications 
from the contributions of the loop strings or non-negligible correlations 
between different string segments \cite{Yamauchi:2010ms}. 
Though these effects are expected to be small, they would
certainly enhance the B-mode or curl-mode signals, and the expected
signal-to-noise ratios will be increased. In this respect, the
analysis based on the simple model may give a rather conservative
estimate for the detectability of cosmic strings. Nevertheless, we
should keep in mind the possibilities that the contributions of loop
strings or non-vanishing correlations are accompanied with changes to
the other components of string stress-energy, together with the
generation of tensor perturbations, which might eventually lead to the
reduction of the B-mode shear and the curl-mode.

\subsection{Angular power spectra and signal-to-noise ratios} 
\label{sec:Observational implications}

\subsubsection{B-mode cosmic shear}
\label{sec:B-mode of cosmic shear fields}

Let us first compute the B-mode power spectrum of cosmic shear field. 
To discuss the weak lensing measurement from imaging surveys, 
we assume the redshift distribution of galaxies, $N$, as 
(e.g., \cite{arXiv:1103.1118,arXiv:1009.3204})
\beq
	N (\cS )\,\dd\cS
		=N_\rmg\frac{3z_\rmS^2}{2\,(0.64z_\rmm )^3}
			\exp\Biggl[ -\left(\frac{z_\rmS}{0.64z_\rmm}\right)^{3/2}\,\Biggr]\,\dd z_\rmS
	\,,\label{eq:background galaxy distribtuion}
\eeq
where $z_\rmm$ and $N_\rmg$ denote the mean redshift and 
the total number of galaxies per square arcminute, respectively.
Taking account of the redshift distribution of galaxies, 
we recast the formula for B-mode power spectra: 
\al{ 
	C_\ell^{\rmB\rmB}
		=&\frac{1}{2\pi}
			\frac{(\ell -1)!(\ell +2)!}{(\ell +1)!(\ell -2)!}
			\int^\infty_0 k^2\dd k\int^\infty_0\dd\chi\int^\infty_0\dd\chi'
	\notag\\
	&\quad\quad\times
			W_1(\chi )\frac{j_\ell (k\chi )}{\chi}W_1(\chi' )\frac{j_\ell (k\chi' )}{\chi'}
			P_{\sigma_\rmg\sigma_\rmg}\left( k;\eta_0 -\chi ,\eta_0 -\chi'\right)
	\,,
        \label{eq:B_mode_spectrum2}
}
where we have introduced the weight function $W_1$, which is the 
normalized distribution function for galaxies along a line-of-sight: 
\al{
	W_1(\chi )=\int^\infty_\chi\dd\cS\,\frac{N(\cS )}{N_\rmg}
	\,.
}
In the cosmic shear measurement, apart from systematics, 
the main noise contribution would come from the intrinsic ellipticity of 
galaxies, which is described by
\beq
	N_\ell^{\rm BB}
		=\frac{\ave{\gamma_{\rm int}^2}}{\hat N}
	\,,
\eeq
where $\ave{\gamma_{\rm int}^2}^{1/2}$ is the root-mean-square intrinsic 
ellipticity, and $\hat N$ is the number density of galaxies per steradians. 
We adopt the empirically derived value, 
$\ave{\gamma_{\rm int}^2}^{1/2}=0.3$~\cite{astro-ph/0107431}, and parameterize 
the number density of galaxies as 
$\hat N=3600N_\rmg (180/\pi )^2\,[{\rm str}^{-1}]$. 
Then, the statistical error of the B-mode power spectrum 
is estimated as~\cite{Dodelson:2003bv,Sarkar:2008ii}
\beq
	\Delta C_\ell^{\rm BB}
		=\sqrt{\frac{2}{(2\ell +1)f_{\rm sky}\Delta\ell}}
			\Bigl( C_\ell^{\rm BB}+N_\ell^{\rm BB}\Bigr)
	\,,
\eeq
where $\Delta\ell$ is the size of multipole bin. For illustrative purpose 
to show the angular power spectra, 
we set $\Delta\ell =(i+1)^3-i^3$ for $i$-th multipole bin. 
To discuss the detectability of the weak lensing signals from cosmic strings, 
we quantify the signal-to-noise ratio (S/N) for the angular power spectrum.  
For the B-mode measurement, it is defined by 
\beq
	\left(\frac{S}{N}\right)^{\rm BB}_{<\ell}
		=\left[\,\sum_{\ell' =2}^\ell
			\left(
				\frac{C^{\rm BB}_{\ell'}}{\Delta C_{\ell'}^{\rm BB}}\right)^2
			\right]^{1/2}
	\,.
\eeq
Note that the signal-to-noise ratio does not depend on the size of 
multipole bins, but depend sensitively on the string parameters,  
$P$ and $G\mu$. To see the detectability of cosmic strings 
through upcoming weak lensing measurements, we consider the three 
representative surveys; HSC, DES, and LSST. Table \ref{survey design} 
summarize the basic parameters for the survey designs.

\begin{figure}[tbp]
\bc
\includegraphics[width=140mm]{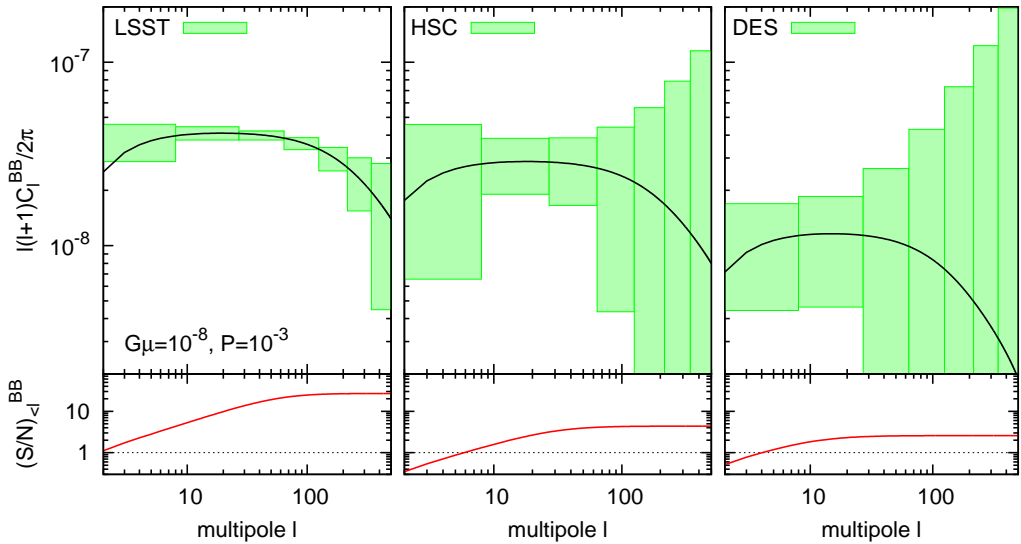}
\caption{
The angular power spectra of the B-mode cosmic shear
from the vector perturbations generated by the cosmic string network 
with $G\mu =10^{-8}$, $P=10^{-3}$
for LSST (left panel), HSC (center panel), and DES (right panel).
The error boxes in each figure show the expected variance of angular power
spectrum from each experiments.
The bottom panels show the signal-to-noise ratio as a function of maximum multipole.
}
\label{BB_SN}
\ec
\end{figure}

\begin{figure}[tbp]
\bc
\includegraphics[width=100mm]{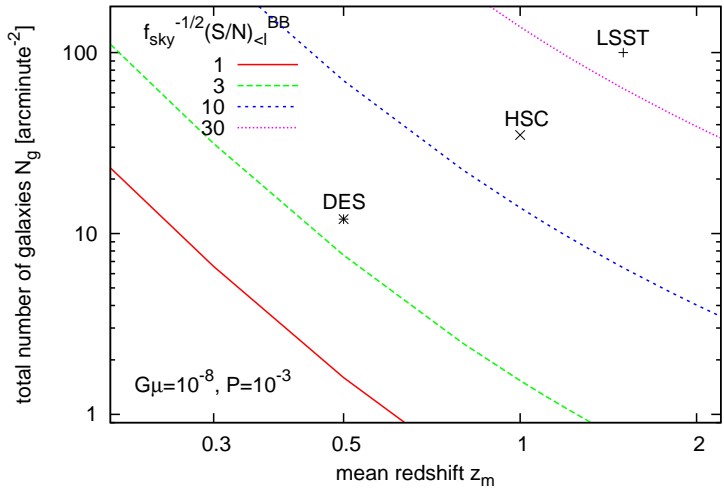}
\caption{
The contour of the signal-to-noise ratio, $f_{\rm sky}^{-1/2}(S/N)_{<200}^{\rm BB}$, 
as the function of $z_\rmm$ and $N_\rmg$.
For the string components, we take $G\mu =10^{-8}$, $P=10^{-3}$\,.
}
\label{BB_SN_zm-Ng_contour}
\ec
\end{figure}

Top panels in Fig.~\ref{BB_SN} show the angular power spectra 
for B-mode cosmic shear induced by a cosmic string network. Here, 
we set the fiducial values of the string parameters to 
$G\mu=10^{-8}$ and $P=10^{-3}$. These fiducial values 
are still consistent with the 
small-scale CMB measurements 
via the GKS effect~\cite{Yamauchi:2010ms}. 
Typically, the B-mode spectrum has a large power with a flat 
shape at large angular scales $\ell\lesssim100$, and it rapidly 
falls off at small angular scales. These features are 
irrespective of the survey design, and are 
determined by the properties of cosmic string network and the 
lensing kernel [eqs.~(\ref{eq:shear EE BB power spectrum}) and 
(\ref{eq:B transfer function}) or eq.~(\ref{eq:B_mode_spectrum2})]. 
On the other hand, the expected amplitude of the power spectrum depends 
not only on string parameters but also on the survey depth ($z_{\rm m}$), 
and the resultant signal-to-noise ratio 
for B-mode spectrum is rather sensitive to the survey specification. 
Each panel of Fig.~\ref{BB_SN} shows the 
expected errors (top) and signal-to-noise ratios (bottom) 
for three representative surveys. As a result, 
a wide and deep survey with dense sampling by LSST is 
capable of detecting the cosmic strings 
with high signal-to-noise ratio $S/N\sim30$. On the other hand, 
comparison of the results 
between HSC and DES indicate that in the cases with a limited 
sky coverage, a deep imaging survey has an advantage to detect the B-mode 
signal with high statistical significance. To see this clearly, we 
allow to vary the survey parameters $(z_{\rm m},\,N_{\rm g})$, and the 
signal-to-noise ratio normalized by the sky coverage, 
$f_{\rm sky}^{-1/2}(S/N)_{<\ell}^{\rm BB}$, is estimated. The results are shown in 
Fig.~\ref{BB_SN_zm-Ng_contour}, where we set 
the maximum multipole to $\ell=200$. 
From this, we roughly estimate the dependence
of survey design as 
$f_{\rm sky}^{-1/2}(S/N)_{<\ell}^{\rm BB}\propto z_{\rm m}^{1.7}N_\rmg^{0.55}$\,.

\begin{figure}[tbp]
\bc
\includegraphics[width=140mm]{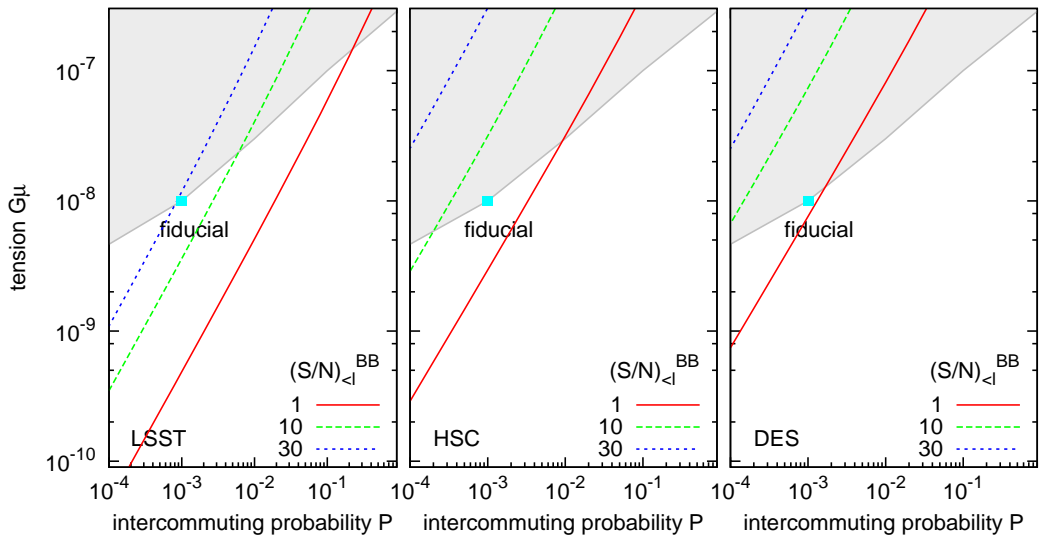}
\caption{
The contours of the signal-to-noise ratio, $(S/N)_{<200}^{\rm BB}$, 
as the function of the tension $G\mu$
and the intercommuting probability $P$ for LSST (left panel), HSC (center panel), and DES (right panel).
The shaded region is excluded from the GKS effect~\cite{Yamauchi:2010ms}.
}
\label{BB_SN_contour}
\ec
\end{figure}

Fig.~\ref{BB_SN_contour} shows 
the potential impact of the weak lensing surveys on the search for 
cosmic string network. Here, varying the cosmic string parameters 
while keeping the fiducial survey setup in 
Table \ref{survey design}, we plot the $(S/N)_{<200}^{\rm BB}$ as function of 
string tension $G\mu$ and intercommuting probability $P$. The resultant 
signal-to-noise ratio is rather sensitive to these two parameters, and 
the detectability would be enhanced for smaller $P$ and larger $G\mu$. 
This is mainly because the power spectrum of vector perturbation 
$P_{\sigma_\rmg\sigma_\rmg}$ roughly scales as 
$P_{\sigma_\rmg\sigma_\rmg}\propto (G\mu )^2\xi^{-5}\propto (G\mu )^2P^{-5/2}$, which 
reflects the fact that a small intercommuting probability increases 
the number density of string segments, and thereby the 
correlation length for string network is reduced. Note that through 
the GKS effect, the small-scale power of CMB temperature anisotropies 
is induced by the cosmic strings, and it roughly scales as 
$\propto (G\mu )^2P^{-3/2}$~\cite{Yamauchi:2010ms}.
The different parameter dependence on the CMB temperature anisotropies 
can be understood as follows : 
the CMB temperature discontinuity induced by strings basically 
has a weak dependence on the length of the string segment $\xi$, and the 
power spectrum amplitude is mostly determined by the number density of the string 
network~\cite{Yamauchi:2010ms}, which implicitly depends on $\xi$. 
As a result, the angular power spectrum for the GKS effect scales as
$\propto (G\mu )^2n_{\rm s}\propto (G\mu )^2\xi^{-3}\propto (G\mu )^2P^{-3/2}$\,,
leading to the different dependence on the intercommuting probability.
Though the results shown here come from the specific model of a string network,
these scalings are expected to be generic and would remain the same.
The region covered by shade in each panel of Fig.~\ref{BB_SN_contour} 
represents the parameters disfavored by the small-scale CMB measurements, 
which are obtained from 
the condition that the string-induced temperature anisotropies 
cannot exceed the measured power spectrum. The 
lower boundary of the shaded region indeed comes from 
the scaling $\propto (G\mu )^2P^{-3/2}$. 
These different behaviors suggest that the weak lensing measurement
can be a complementary probe of the cosmic strings, and is
advantageous for detecting a string network with small intercommuting
probability. That is, the combined analysis of the weak lensing and
small-scale CMB measurements would be quite essential not only to
obtain a tight constraint on string parameters, but also to break the
parameter degeneracies. The precision measurement of the large angle
CMB temperature anisotropies would be also helpful to obtain a tighter
constraint on the parameters. We hope to come back these issues in a
future publication.
Fig.~\ref{BB_SN_contour} implies that for $P\lesssim10^{-1}$,  
B-mode signal of cosmic strings is detectable for a small 
string tention $G\mu\sim5\times10^{-8}$. For theoretically inferred 
smallest value $P\sim10^{-3}$, we could even detect the signal for 
$G\mu\sim5\times10^{-10}$.


\bc
\begin{table}[tbp]
\bc
\begin{tabular}{cccc} \hline 
Survey & $f\rom{sky}$ & $z_\rmm$ & $N_\rmg$ [$\rm{arcmin}^{-2}$]\\ 
\hline
\hline
HSC~\cite{HSC} & 0.05 (2000\,${\rm deg}^2$) & 1.0 & 35 \\
\hline 
DES~\cite{astro-ph/0510346} & 0.125 (5000\, ${\rm deg}^2$) & 0.5 & 12 \\ 
\hline 
LSST~\cite{arXiv:0912.0201} & 0.5 (20000\, ${\rm deg}^2$) & 1.5 & 100 \\ 
\hline
\hline 
\end{tabular}
\caption{
Survey design for HSC, DES, and LSST.
The sky coverage $f_{\rm sky}$, the mean redshift $z_\rmm$,
and the number of the galaxies per square arcminute $N_\rmg$
are shown.}\label{survey design}
\ec
\end{table} 
\ec

\bc
\ec

\bc
\ec

\bc
\ec

\subsubsection{Curl-mode deflection angle}
\label{sec:Curl-mode of deflection angle}

Let us next consider the curl-mode signals from the CMB measurements. 
We calculate the expected curl-mode signal and 
the signal-to-noise ratio for the upcoming and idealistic CMB experiments.  
We consider the combination of 
small- and large-scale CMB measurements by ACTPol and PLANCK (ACTPol+PLANCK) 
for a representative upcoming/on-going experiment, 
and the high-resolution full-sky experiment limited by the cosmic variance,  
just for illustrative purpose. 
We assume that the curl-mode deflection angle is reconstructed from the 
lensed CMB map based on the quadratic reconstruction technique~\cite{Zaldarriaga:1998te,Hu:2001tn,Hu:2001kj,Okamoto:2003zw}.
Similar to the B-mode cosmic shear, we define the signal-to-noise ratio
for curl-mode deflection angle: 
\beq
	\left(\frac{S}{N}\right)^{\varpi\varpi}_{<\ell}
		=\left[\,\sum_{\ell' =2}^\ell
			\left(
				\frac{C^{\varpi\varpi}_{\ell'}}{\Delta C_{\ell'}^{\varpi\varpi}}\right)^2
			\right]^{1/2}
	\,,
\eeq
with the error $\Delta C_\ell^{\varpi\varpi}$ given by 
\beq
	\Delta C_\ell^{\varpi\varpi}
		=\sqrt{\frac{2}{(2\ell +1)f_{\rm sky}\Delta\ell}}
			\Bigl( C_\ell^{\varpi\varpi}+N_\ell^{\varpi ,(c)}\Bigr)
	\,.
\eeq
Here, $N_\ell^{\varpi ,(c)}$ is the reconstruction noise spectrum for
the optimal combination of the quadratic estimator~\cite{Namikawa:2011cs}. 
In Appendix \ref{sec:Reconstruction noise}, the explicit expression for 
the noise spectrum of the curl mode is presented, and 
its dependence on the experimental specification is briefly summarized.

\begin{figure}[tbp]
\bc
\includegraphics[width=120mm]{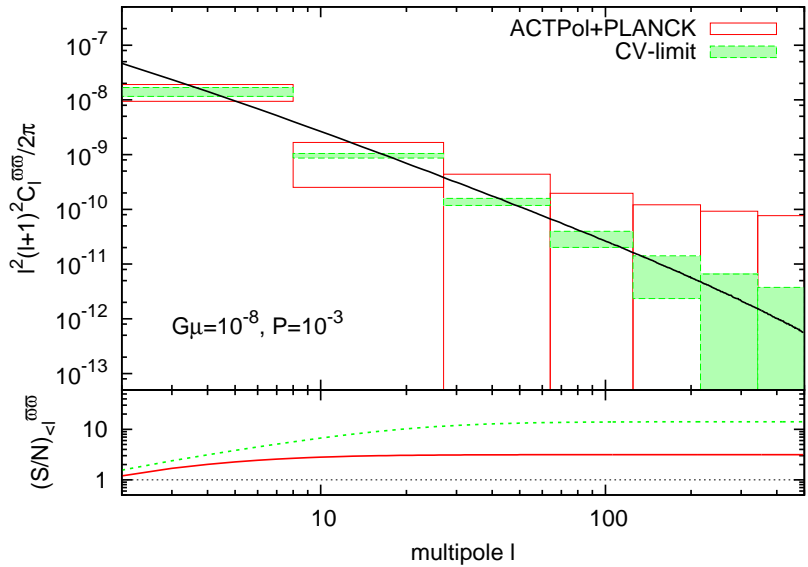}
\caption{
The angular power spectrum of the curl-mode deflection angle
from the vector perturbations generated by the cosmic string network
with $G\mu =10^{-8}$, $P=10^{-3}$. 
The error boxes represent the expected variance of angular power
spectrum from ACTPol+PLANCK (empty red), CV-limit (shaded green), 
with the multipole used in the reconstruction procedure,
$\ell_{\rm max}=7000$ (see \cite{Namikawa:2011cs} and Appendix 
\ref{sec:Reconstruction noise})\,. Just for illustration, 
we set the size of multipole bins 
to $\Delta\ell =(i+1)^3-i^3$ for $i$-th bin. 
The bottom panel shows the signal-to-noise ratio as a function of maximum multipole
for each survey.
}
\label{varpivarpi_SN}
\ec
\end{figure}

\begin{figure}[tbp]
\bc
\includegraphics[width=120mm]{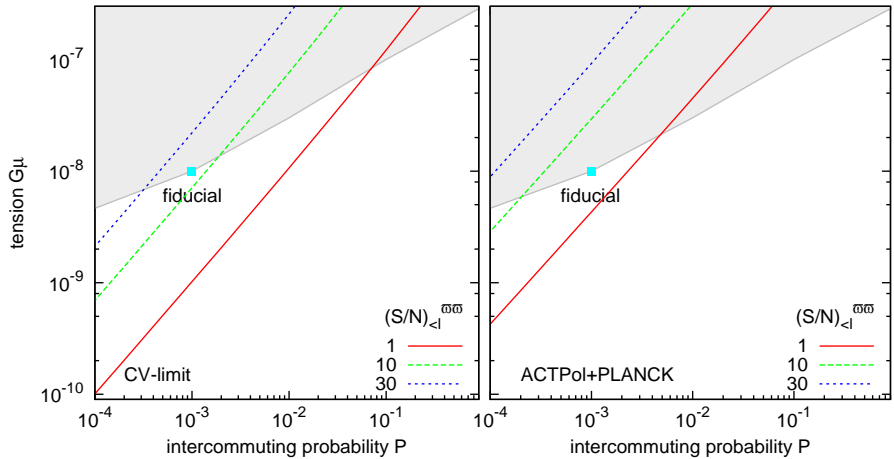}
\caption{
The contours of the signal-to-noise ratio, $(S/N)_{<200}^{\varpi\varpi}$,
as the function of the tension $G\mu$
and the intercommuting probability $P$ 
for CV-limit (left panel) and ACTPol+PLANCK (right panel).
The shaded region is excluded from the GKS effect~\cite{Yamauchi:2010ms}.
}
\label{varpivarpi_SN_contour}
\ec
\end{figure}

Fig.~\ref{varpivarpi_SN} shows the 
expected curl-mode signal for the cosmic strings. 
Top and bottom panels respectively plot 
the angular power spectrum and the signal-to-noise ratio for the 
pseudo-scalar lensing potential, $\varpi$, assuming the 
string parameters $G\mu=10^{-8}$ and $P=10^{-3}$. 
At large-angular scales, the curl-mode signal is 
prominent and has the largest amplitude, but the 
power spectrum rapidly falls off at small scales. 
Considering the fact that a measurement of 
string-induced CMB anisotropies via the 
GKS effect is only available at small scales, 
CMB-lensing experiment can be also a complementary probe, and would 
be more suited for the detection of a cosmic string network. 
Fig.~\ref{varpivarpi_SN} suggests that 
for a definite detection with $S/N\gtrsim10$, 
a full-sky lensing experiment would be ideal and the best, but   
even with the upcoming experiment of ACTPol+PLANCK, 
we can detect the signature of cosmic strings with $S/N\sim3$. 
Recalling that there would be additional contributions to the 
angular power spectrum, 
leading to an enhancement of the power spectrum amplitude,  
the results shown here should be regarded as a rather conservative estimate, 
and the actual detectability might be increased.
Finally, Fig.~\ref{varpivarpi_SN_contour} shows the dependence of signal-to-noise
ratio on the string parameters $G\mu$ and $P$. Similar to the 
B-mode cosmic shear, the signal-to-noise ratio for curl-mode signal scales as
$(S/N)_{<\ell}^{\varpi\varpi}\propto (G\mu )^2P^{-5/2}$, and the CMB-lensing 
experiment is capable of detecting a string network with small $P$. Since 
the small-scale CMB experiment is usually dominated by the contributions 
from point sources and the Sunyaev-Zel'dovich (SZ) effect, 
the curl-mode measurement would provide not only a direct probe of 
cosmic strings, but also a diagnosis helpful to check 
the systematics in the derived constraints from the GKS effect.

\bc
\ec

\bc
\ec

\section{Summary} \label{sec:Summary} 

In this paper, we have discussed the observational signature of 
the vector metric perturbations through the effect of weak gravitational 
lensing.  In the presence of vector perturbations, 
the non-vanishing 
signals for B-mode cosmic shear and curl-mode deflection angle 
naturally appears, and these would be a unique signature of vector 
perturbations. Solving the geodesic and geodesic deviation equations, 
we have derived the full-sky formulas for angular power spectra of weak 
lensing signals, and give the explicit expressions for E-/B-mode cosmic 
shear [expression \eqref{eq:shear EE BB power spectrum} with 
eqs.~\eqref{eq:E transfer function}-\eqref{eq:beta^1 def}], 
and gradient-/curl-mode deflection angle 
[expression \eqref{eq:scalar and pseudo-scalar power spectrum} with 
eqs.~\eqref{eq:phi transfer function} and \eqref{eq:varpi transfer function}].

As a possible source for seeding vector perturbations, 
we then considered a cosmic string network, and discuss
its detectability from upcoming weak lensing and CMB measurements. 
Based on the formulas and a simple model for cosmic string network, 
we calculated the angular power spectra, and 
the expected signal-to-noise ratios for the B-mode cosmic shear 
and curl-mode deflection angle were estimated. 
The string-induced signals typically have a large power at large-angular 
scales, and we found that the signals with small intercommuting 
probability $P$ are detectable from future lensing experiments. 
With the theoretically inferred smallest value $P\sim10^{-3}$, we 
could even detect the cosmic strings with $G\mu \sim5\times10^{-10}$. 
Therefore, the weak lensing measurement of the B-mode cosmic shear and 
curl-mode deflection angle would be 
an important probe for cosmic string network, and is complementary to 
the small-scale CMB experiment via the GKS effect.

Throughout the paper, we have assumed several idealizations;   
the weak lensing measurement are perfect without annoying masking 
effect, and free from the foreground contaminations. 
In practice, the B-mode cosmic shear would be contaminated 
by the E/B-mode mixing arising from the incomplete sky coverage~\cite{Hikage:2010sq}, 
intrinsic alignment of galaxy images induced by the gravitational 
clustering~\cite{Heymans:2002uw}, and the shear-intrinsic ellipticity
correlations~\cite{Hirata:2007np,Mandelbaum:2005hr}. 
Also, the curl-mode deflection angle from the CMB maps 
is affected not only by 
the point sources and SZ effect \cite{Das:2011ak,Amblard:2004ih}, 
but also by the masking effect~\cite{Carvalho:2011gx,Bucher:2010iv}
and the inhomogeneous noises~\cite{Hanson:2009dr}.
As for the detection of cosmic string network, we have discussed 
the expected weak lensing signals based on a very simple model, 
and the model prediction should be further improved for 
future application to the lensing measurements. 
There are several missing pieces, including the contribution of 
tensor perturbations \cite{Damour:2000wa,Damour:2001bk,
Damour:2004kw,Kuroyanagi:2012wm,Kawasaki:2010yi} and 
the correlation between different string segments  
\cite{Yamauchi:2010ms}, which may enhance the signal-to-noise ratio.   
We hope to come back these issues near future.


\acknowledgments
We would like to thank Fabian Schmidt and Donghui Jeong for crucial comments.
This work is supported in part by a Grant-in-Aid for Scientific Research from 
JSPS (No. 24540257). 
\appendix

\section{Useful formula} 

We summarize the formulas used in this paper.

\subsection{Spherical Bessel function} 
\label{sec:Spherical Bessel function}

The spherical Bessel functions, $j_\ell (x)$, 
are solutions to the differential equation:
\al{
	j_\ell'' (x)+\frac{2}{x}j_\ell' (x)
		+\left( 1-\frac{\ell (\ell +1)}{x^2}\right) j_\ell (x)
		=0
	\,.\label{eq:spherical Bessel function eq}
}
The recursion relations of spherical Bessel functions are given by
\al{
	&\frac{j_\ell (x)}{x}
		=\frac{1}{2\ell +1}\Bigl\{ j_{\ell -1}(x)+j_{\ell +1}(x)\Bigr\}
	\,,\\
	&j_\ell' (x)
		=\frac{1}{2\ell +1}\Bigr\{\ell j_{\ell -1}(x)+(\ell +1)j_{\ell +1}(x)\Bigr\}
		=\ell\,\frac{j_\ell (x)}{x}-j_{\ell +1}(x)
	\,.\label{eq:Bessel function recursion relation}
}

\subsection{Legendre polynomials} 

The associated Legendre functions are defined in terms of the Legendre polynomials by
\al{
	P_{\ell ,m}(\mu )=(-1)^m(1-\mu^2 )^{m/2}\frac{\dd^m}{\dd\mu^m}P_\ell (\mu )
	\,,
}
where the Legendre polynomials are solutions to the differential equation:
\al{
	\frac{\dd}{\dd\mu}\left( (1-\mu^2 )\frac{\dd}{\dd\mu}P_\ell (\mu )\right)
		+\ell (\ell +1)P_\ell (\mu )=0
	\,.
}
We then take the integration of the associated Legendre functions:
\al{
	\int^1_{-1}\dd\mu\sqrt{1-\mu^2}&P_{\ell ,+1}(\mu )e^{-ix\mu}
		=-\int^1_{-1}\dd\mu\left( 1-\mu^2\right) e^{-ix\mu}\frac{\dd}{\dd\mu}P_\ell (\mu )
	\notag\\
		=&-i\Bigl( 2\pd_x +x \left( 1+\pd_x^2\right)\Bigr) \int^1_{-1}\dd\mu\, e^{-ix\mu}P_\ell (\mu )
	\notag\\
		=&2(-i)^{\ell +1}\Bigl( 2\pd_x +x(1+\pd_x^2 )\Bigr) j_\ell (x)
	\notag\\
		=&2(-i)^{\ell +1}\frac{(\ell +1)!}{(\ell -1)!}\frac{j_\ell (x)}{x}
	\,,\label{eq:angular integration of P_l1}
}
where we have used the differential equation for the spherical Bessel function
eq.~\eqref{eq:spherical Bessel function eq}.

\subsection{Spherical harmonics} 

A scalar field on the sphere can be expanded in (spin-$0$) spherical
harmonics, $Y_{\ell m}(\hatn )$\,. 
Spherical harmonics can be written in terms of the Legendre polynomials as
\beq
	Y_{\ell m}(\hatn )
		=\sqrt{\frac{2\ell +1}{4\pi}\frac{(\ell -m)!}{(\ell +m)!}}
			P_{\ell ,m}(\mu )e^{im\varphi}
	\,,
\eeq
where $\mu =\cos\theta$\,.
Then one can verify the angular integration of the spherical harmonics:
\al{
	\int\dd^2\hatn Y_{\ell m}^*(\hatn )e^{\pm i\varphi}e^{-ix\mu}
		=&\int^1_{-1}\dd\mu\sqrt{1-\mu^2}
			\int^{2\pi}_0\dd\varphi\, Y^*_{\ell m}(\hatn )e^{\pm i\varphi}e^{-ix\mu}
	\notag\\
		=&\pm 2\pi\,\delta_{m,\pm 1}\sqrt{\frac{2\ell +1}{4\pi}\frac{(\ell -1)!}{(\ell +1)!}}
			\int^1_{-1}\dd\mu\sqrt{1-\mu^2}\,P_{\ell ,+1}(\mu )e^{-ix\mu}
	\notag\\
		=&\pm (-i)^{\ell +1}\delta_{m,\pm 1}\sqrt{4\pi (2\ell +1)\frac{(\ell +1)!}{(\ell -1)!}}\,
			\frac{j_\ell (x)}{x}
	\,,\label{eq:angular integration of Y_lm}
}
where we have used the angular integration of the Legendre polynomials, eq.~\eqref{eq:angular integration of P_l1}.

\section{Christoffel symbols and Riemann tensors}
\label{sec:Christoffel symbols and Riemann tensors from vector perturbations}

The Christoffel symbols on the unperturbed spacetime, namely Minkowski spacetime, 
in the Cartesian coordinate system are trivially $\Gamma^\rho_{\mu\nu}=0$\,.
Since the Christoffel symbols are not covariant quantities, 
the unperturbed Christoffel symbols in the spherical coordinate system 
can have the components:
\al{
	\Gamma^\chi_{ab}=-\chi\,\omega_{ab}
	\,,\ \ 
	\Gamma^a_{\chi b}=\frac{1}{\chi}\,\delta^a{}_b
	\,,\ \ 
	\Gamma^a_{bc}=^{(2)}\Gamma^a_{bc}
	\,,\ \ 
	\text{otherwise}=0
	\,,
}
where $a,b,c=\theta ,\varphi$ and the two-dimensional Christoffel symbols are
\beq
	^{(2)}\Gamma^\theta_{\varphi\varphi}=-\sin\theta\cos\theta
	\,,\ \ 
	^{(2)}\Gamma^\varphi_{\theta\varphi}=\cot\theta
	\,,\ \ 
	\text{otherwise}=0
	\,.
\eeq
We can calculate the linearized Christoffel symbols as 
$\delta\Gamma^\rho_{\mu\nu}
=\frac{1}{2}g^{\rho\sigma}\left( h_{\sigma\mu ;\nu}+h_{\sigma\nu ;\mu}-h_{\mu\nu ;\sigma}\right)$\,.
Hence we have the components of the linearized Christoffel symbols
induced by the vector perturbations, $h_{0i}=-\sigma_{\rmg ,i}$\,, as
\beq
	\delta\Gamma^i_{00}=-\dot\sigma_\rmg{}^i
	\,,\ \ 
	\delta\Gamma^i_{0j}=\frac{1}{2}\left(\sigma_{\rmg ,j}{}^{|i}-{\sigma_\rmg{}^i}\,{}_{|j}\right)
	\,,\ \ 
	\delta\Gamma^0_{ij}=\frac{1}{2}\left(\sigma_{\rmg ,i|j}+\sigma_{\rmg ,j|i}\right)
	\,,\ \ 
	\text{otherwise}=0
	\,.
\eeq
Recalling that the geodesic in the background spacetime can be solved as 
$x^\mu (\chi )=(\eta_0 -\chi ,\chi\, e^i_\chi )$\,, we can calculate
\beq
	\delta\Gamma^i_{\mu\nu}\frac{\dd x^\mu}{\dd\chi}\frac{\dd x^\nu}{\dd\chi}
		=\delta\Gamma^i_{00}-2\delta\Gamma^i_{0j}e^j_\chi
			+\delta\Gamma^i_{jk}e^j_\chi\, e^k_\chi
		=-\dot\sigma_\rmg{}^i+{\sigma_\rmg{}^i}{}_{|k}e^k_\chi
			-\sigma_{\rmg ,j}{}^{|i}e^j_\chi
	\,.\label{eq:deltaGamma}
\eeq
Multiplying $e_i^a$ in both side of eq.~\eqref{eq:deltaGamma} and 
using the definition of the covariant derivative associated with $\omega_{ab}$, 
eqs.~\eqref{eq:basis derivative}, \eqref{eq:intrinsic covariant derivative}, 
we obtain the angular component of eq.~\eqref{eq:deltaGamma}:
\beq
	e_i^a\delta\Gamma^i_{\mu\nu}\frac{\dd x^\mu}{\dd\chi}\frac{\dd x^\nu}{\dd\chi}
		=\frac{1}{\chi}\omega^{ab}
			\biggl\{
				\frac{\dd}{\dd\chi}\left(\chi\,\sigma_{\rmg ,b}\right)
				-\left(_0\sigma_\rmg\right)_{:b}
			\biggr\}
	\,,\label{eq:perturbed Christoffel symbols}
\eeq
where we have introduced $\dd /\dd\chi =e^i_\chi\pd_i -\pd_\eta$\,,
$_0\sigma_\rmg =\sigma_{\rmg ,i}e^i_\chi$\,,
$\sigma_{\rmg ,a}=\sigma_{\rmg ,i}e^i_a$\,.

Since the background geometry is Minkowski spacetime, 
the Riemann tensor and the symmetric optical tidal matrix 
at the zeroth-order in metric perturbations 
are trivially given as $R_{\mu\rho\nu\sigma}(g_{\alpha\beta})=0$ 
and $\bar\mcT^a{}_b(g_{\alpha\beta})=0$\,. 
Using the explicit expression for the linearized Riemann tensor as 
$\delta R_{\mu\rho\nu\sigma}=\frac{1}{2}\left( -h_{\mu\nu ;\rho\sigma}-h_{\rho\sigma ;\mu\nu}
+h_{\mu\sigma ;\rho\nu}+h_{\nu\rho ;\mu\sigma}\right)$\,, 
we have
\al{
	&\delta R_{0i0j}
		=-\frac{1}{2}\left(\dot\sigma_{\rmg ,i|j}+\dot\sigma_{\rmg ,j|i}\right)
	\,,\ \ 
	\delta R_{ik0j}=\frac{1}{2}\left(\sigma_{\rmg ,i|kj}-\sigma_{\rmg ,k|ij}\right)
	\,,\ \ 
	\delta R_{ijkl}=0
	\,.
}
With a help of eqs.~\eqref{eq:basis derivative}, \eqref{eq:intrinsic covariant derivative}, 
we can calculate the linearized symmetric optical tidal matrix as
\al{
	\delta\mcT_{ab}
		=&-\frac{1}{E^2}\delta R_{\mu\rho\nu\sigma}
			k^\mu k^\nu e^\rho_a e^\sigma_b
		=-\delta R_{0i0j}e^i_{(a}e^j_{b)}
			+2\,\delta R_{ki0j}e^k_\chi e^i_{(a}e^j_{b)}
	\notag\\
		=&e^k_\chi e^i_{(a}e^j_{b)}\sigma_{\rmg ,k|ij}
			-\frac{\dd}{\dd\chi}\left( e^i_{(a}e^j_{b)}\sigma_{\rmg ,i|j}\right)
	\notag\\
		=&\,\frac{1}{\chi^2}
			\left\{
				\left(_0\sigma_\rmg\right)_{:ab}
				-\frac{\dd}{\dd\chi}\left(\chi\,\sigma_{\rmg ,(a:b)}\right)
				+\chi\,\omega_{ab}\left({}_0\dot\sigma_\rmg\right)
			\right\}
	\,.\label{eq:linearized optical tidal matrix}
}

\section{Derivation of angular power spectrum} 

In this section, we provide details of the derivation of the full-sky formulas 
for the angular power spectrum for the E-/B-mode cosmic shear and
the gradient-/curl-mode deflection angle generated by
the vector perturbations.
We first define the spin-raising/lowering operators, spin-weighted spherical harmonics and 
present the explicit relation between the intrinsic covariant derivative
on the unit sphere and the spin-raising/lowering operators 
in section \ref{sec:Spin operators and spin-weighted spherical harmonics}.
In section \ref{sec:E-/B-modes of shear fields} and \ref{sec:Scalar/pseudo-scalar lensing potential}
we present the formula for the angular power spectrum for 
the E-/B-mode cosmic shear and the gradient-/curl-mode deflection angle
generated by the vector perturbations, following \cite{Zaldarriaga:1996xe,Lin:2004xy}.
In section \ref{sec:Derivation of shear-deflection relation} we derive
the explicit relation between the cosmic shear field and the deflection angle.

\subsection{Spin operators and spin-weighted spherical harmonics}
\label{sec:Spin operators and spin-weighted spherical harmonics}

We define a pair of operator $\pspin$ and $\mspin$,
known as spin-raising and lowering operators, respectively.
These operators have the properties of
increasing or decreasing the index of the spins by $1$.
For a spin-$s$ function $_sX$\,, their explicit forms are~\cite{Zaldarriaga:1996xe,Lin:2004xy}
\al{
	\pspin\left(_s X\right)
		\equiv&-\sin^s\theta
			\left(\pd_\theta +\frac{i}{\sin\theta}\pd_\varphi\right)
			\sin^{-s}\theta\,\left(_s X\right)
	\,,\label{eq:pspin def}\\
	\mspin\left(_sX\right)
		\equiv&-\sin^{-s}\theta
			\left(\pd_\theta -\frac{i}{\sin\theta}\pd_\varphi\right)
			\sin^s\theta\,\left({}_sX\right)
    \,.\label{eq:mspin def}
} 
Since we are interested in the spin-$\pm 2$ quantities, 
acting twice with the spin-raising/lowering operators on the spin-$\pm 2$ fields, 
${}_{\pm 2}X$\,, gives
\al{
	&\mspin^2\,\left(_{+2}X\right)
			=\left( -\pd_\mu -\frac{i}{1-\mu^2}\pd_\varphi\right)^2
				\Bigl[
					\left( 1-\mu^2\right)\,\left(_{+2}X\right)
				\Bigr]
	\,,\label{eq:mspin^2 spin+2X}\\
	&\pspin^2\,\left(_{-2}X\right)
			=\left( -\pd_\mu +\frac{i}{1-\mu^{2}}\pd_\varphi\right)^2
				\Bigl[
					\left( 1-\mu^2\right)\,\left(_{-2}X\right)
				\Bigr]
	\,.\label{eq:pspin^2 spin-2X}
}
where we have used the directional cosine $\mu =\cos\theta$ and $\pd_\mu =\pd /\pd\mu$\,.
To see the relation between the intrinsic covariant derivatives on the unit sphere
and the spin-raising/lowering operators, one can verify 
\al{
	\chi\, e^j_\pm\pd_j e^i_\pm
		=\cot\theta\, e^i_\pm
	\,,\ \ 
	\chi\, e^j_\pm\pd_j e^i_\mp
		=-2\, e^i_\chi -\cot\theta\, e^i_\mp
	\,.
} 
where we have used eq.~\eqref{eq:basis derivative}\,.
In terms of the spin basis, these are reduced to
\al{
	e^a_{\pm\, :b}\,e^b_\pm =\cot\theta\, e^a_\pm
	\,,\ \ 
	e^a_{\pm\, :b}\,e^b_\mp =-\cot\theta\, e^a_\pm
	\,.
} 
A spin-$s$ function, $_sX$\,, can be written in terms of the spin basis and
a symmetric trace-free rank-$s$ tensor, $X_{a_1\cdots a_s}$, as
\al{
	_sX=X_{a_1\cdots a_s}e^{a_1}_+\cdots e^{a_s}_+
	\,,\ (s\geq 0)
	\,,\ \ \ 
	{}_sX=X_{a_1\cdots a_{|s|}}e^{a_1}_-\cdots e^{a_{|s|}}_-
	\,,\ (s<0)
	\,,
} 
with $e^a_\pm\equiv e^a_ie^i_\pm$\,.
With these notations, we can easily prove the following useful relations:
\al{
	&\left(_0 X\right)_{:a}\,e^a_+=-\pspin\,\left(_0 X\right)
	\,,\ \ 
	\left(_0 X\right)_{:a}\,e^a_-=-\mspin\, \left(_0 X\right)
	\,,\label{eq:spin op1}\\
	&X_{a:b}\,e^a_+e^b_+=-\pspin\, \left(_{+1}X\right)
	\,,\ \ 
	X_{a:b}\,e^a_-e^b_-=-\mspin\, \left(_{-1}X\right)
	\,,\label{eq:spin op2}\\
	&X_{a:b}\,e^a_-e^b_+=-\pspin\, \left(_{-1}X\right)
	\,,\ \ 
	X_{a:b}\,e^a_+e^b_-=-\mspin\, \left(_{+1}X\right)
	\,,\label{eq:spin op3}\\
	&\left(_0 X\right)_{:ab}\,e^a_+e^b_+=\pspin^2\, \left(_0 X\right)
	\,,\ \ 
	\left(_0 X\right)_{:ab}\,e^a_-e^b_-=\mspin^2\, \left(_0 X\right)
	\,,\label{eq:spin op4}\\
	&\left(_0 X\right)_{:ab}\,e^a_+e^b_-
		=\pspin\mspin\, \left(_0 X\right)
		=\mspin\pspin\, \left(_0 X\right)
	\,.\label{eq:spin op5}
} 

With the spin operators, one can express the spin-weighted spherical harmonics, 
$_sY_{\ell m}$\,, in terms of the spin-$0$ spherical harmonics, $Y_{\ell m}$ as
\al{
	_sY_{\ell m}(\hatn )
		=\sqrt{\frac{(\ell -s)!}{(\ell +s)!}}\,
			\pspin^s\, Y_{\ell m}(\hatn )
	\,,\label{eq:spin-weighted positive spherical harmonics}
} 
for $0\leq s\leq\ell$\,, and
\al{
	_sY_{\ell m}(\hatn )
		=\sqrt{\frac{(\ell +s)!}{(\ell -s)!}}\,
			\mspin^{-s}\, Y_{\ell m}(\hatn )
	\,,\label{eq:spin-weighted negative spherical harmonics}
} 
for $-\ell\leq s\leq 0$\,.
One can also see the following useful properties of the spin-weighted
spherical harmonics:
\al{
	&\pspin\,\Bigl({}_s Y_{\ell m}(\hatn )\Bigr)
		=\sqrt{(\ell -s)(\ell +s+1)}\,\,_{s+1}Y_{\ell m}(\hatn )
	\,,\label{eq:pspin _sY_lm}\\
	&\mspin\,\Bigl({}_s Y_{\ell m}(\hatn )\Bigr)
		=-\sqrt{(\ell +s)(\ell -s+1)}\,\,_{s-1}Y_{\ell m}(\hatn )
	\,.\label{eq:mspin _sY_lm}
} 

\subsection{E-/B-mode cosmic shear} 
\label{sec:E-/B-modes of shear fields}

Since we are interested in the spin-$\pm 2$ quantities, $g$ and $g^*$\,,
it is convenient to introduce the spin-$0$ quantities, which are 
constructed from the spin-$\pm 2$ quantities.
For real space calculations it is useful to introduce the spin-$0$ quantities, 
$\mspin^2\, g (\hatn )$ and $\pspin^2\, g^*(\hatn )$~\cite{Zaldarriaga:1996xe,Lin:2004xy}:
\al{
	&\mspin^2 g(\hatn )
		=\sum_{\ell =2}^\infty\sum_{m=-\ell}^\ell
			{}_{+2}g_{\ell m}\,\,\mspin^2\Bigl({}_{+2}Y_{\ell m}(\hatn )\Bigr)
		=\sum_{\ell =2}^\infty\sum_{m=-\ell}^\ell
			\sqrt{\frac{(\ell +2)!}{(\ell -2)!}}\,\,_{+2}g_{\ell m}\, Y_{\ell m}(\hatn )
	\,,\label{eq:mspin^2 g}\\
	&\pspin^2 g^*(\hatn )
		=\sum_{\ell =2}^\infty\sum_{m=-\ell}^\ell
			{}_{-2}g_{\ell m}\,\,\pspin^2\Bigl({}_{-2}Y_{\ell m}(\hatn )\Bigr)
		=\sum_{\ell =2}^\infty\sum_{m=-\ell}^\ell
			\sqrt{\frac{(\ell +2)!}{(\ell -2)!}}\,\,_{-2}g_{\ell m}\, Y_{\ell m}(\hatn )
	\,,\label{eq:pspin^2 g^*}
}
where we have used eqs.~\eqref{eq:spherical expansion for reduced shear}\,, 
\eqref{eq:pspin _sY_lm} and \eqref{eq:mspin _sY_lm}.
Furthermore, we can introduce $\tilde g^\rmE (\hatn )$ and $\tilde g^\rmB (\hatn )$\,,
constructed from the spin-$0$ quantities, $\mspin^2 g$ and $\pspin^2 g^*$:
\al{
	&\tilde g^\rmE (\hatn )
		\equiv -\frac{1}{2}\Bigl(\mspin^2 g(\hatn )+\pspin^2 g^*(\hatn )\Bigr)
		\equiv\sum_{\ell =2}^\infty\sum_{m=-\ell}^\ell\,\tilde g_{\ell m}^\rmE\, Y_{\ell m}(\hatn )
	\,,\label{eq:tilde g^E def}\\
	&\tilde g^\rmB (\hatn )
		\equiv -\frac{1}{2i}\Bigl(\mspin^2 g(\hatn )-\pspin^2 g^*(\hatn )\Bigr)
		\equiv\sum_{\ell =2}^\infty\sum_{m=-\ell}^\ell\,\tilde g_{\ell m}^\rmB\, Y_{\ell m}(\hatn )
	\,.\label{eq:tilde g^B def}
}
Combining with eqs.~\eqref{eq:mspin^2 g}-\eqref{eq:tilde g^B def},
one can see the explicit relation between the multipole coefficients 
of the spin-$0$ quantities, $\tilde g^\rmE$ and $\tilde g^\rmB$\,, and of the spin-$\pm 2$
quantities, $g^\rmE$ and $g^\rmB$ (see eq.~\eqref{eq:E-/B-mode def})\,, as
\al{
	&\tilde g^\rmE_{\ell m}
		=-\frac{1}{2}\sqrt{\frac{(\ell +2)!}{(\ell -2)!}}\,
			\Bigl({}_{+2}g_{\ell m}+_{-2}g_{\ell m}\Bigr)
		=\sqrt{\frac{(\ell +2)!}{(\ell -2)!}}\,\,g^\rmE_{\ell m}
	\,,\label{eq:tilde g^E and g^E}\\
	&\tilde g^\rmB_{\ell m}
		=-\frac{1}{2i}\sqrt{\frac{(\ell +2)!}{(\ell -2)!}}\,
			\Bigl({}_{+2}g_{\ell m}-_{-2}g_{\ell m}\Bigr)
		=\sqrt{\frac{(\ell +2)!}{(\ell -2)!}}\,\,g^\rmB_{\ell m}
	\,.\label{eq:tilde g^B and g^B}
}
Once we obtain the multipole coefficients of the spin-$0$ quantities, 
the E-/B-mode cosmic shear can be calculated by using eqs.~\eqref{eq:tilde g^E and g^E}
and \eqref{eq:tilde g^B and g^B}.
\\

Let us now derive the explicit expression for the reduced shear
and E-/B-mode angular power spectra in terms of spin-weighted quantities.
Using eqs.~\eqref{eq:spin op2}, \eqref{eq:spin op4}, we can rewrite 
eq.~\eqref{eq:reduced shear from vector perturbations} 
in terms of the spin operators:
\al{
	g=&-\frac{1}{2}\int^{\cS}_0\frac{\dd\chi}{\chi}\,
			\biggl\{
				\frac{\cS -\chi}{\cS}\,\pspin^2\left(_0\sigma_\rmg\right)
				+\,\pspin\left(_{+1}\sigma_\rmg\right)
			\biggr\}
	\,,\label{eq:reduced shear in spin op}
}
where $_0\sigma_\rmg =\sigma_{\rmg ,i}e^i_\chi$\,, and
$_{\pm 1}\sigma_\rmg =\sigma_{\rmg , i} e^i_\pm$\,.
To compute the angular power spectrum, we fix the coordinate system.
Without loss of generality, we can always choose coordinates such that
$k^i\equiv k\hat k^i=k(0,0,1)$, $e^i_\pm (\hat{\bm k})=(1,\pm i ,0)$\,.
It is useful to introduce the directional cosine 
$\mu =\hat k_i\,e^i_\chi (\hatn )=\cos\theta$ rather than $\theta$\,.
We then evaluate
\al{ 
	&e_{\chi ,i} (\hatn )e^i_\pm (\hat{\bm k})=\sqrt{1-\mu^2}e^{\pm i\varphi}
	\,,\label{eq:direct product1}\\
	&e_{+,i}(\hatn )e^i_\pm (\hat{\bm k})=\left(\mu\mp 1\right) e^{\pm i\varphi}
	\,,\ \ 
	e_{-,i}(\hatn )e^i_\pm (\hat{\bm k})=\left(\mu\pm 1\right) e^{\pm i\varphi}
	\,.\label{eq:direct product2}
} 
Decomposing the vector perturbations into the Fourier modes (see eq.~\eqref{eq:Fourier exp sigma}),
we multiply $\sigma_{\rmg ,i}$ by $e^i_\chi (\hatn )$ and $e^i_\pm (\hatn )$\,.
With a help of eqs.~\eqref{eq:direct product1}, \eqref{eq:direct product2},
we obtain 
\al{ 
	&_0\sigma_\rmg
		\equiv\sigma_{\rmg ,i}e^i_\chi
		=\int\frac{\dd^3{\bm k}}{(2\pi )^3}e^{-ik\chi\mu}\sqrt{1-\mu^2}
			\left(
				\sigma_\rmg^{(+1)}({\bm k},\eta )e^{i\varphi}
				+\sigma_\rmg^{(-1)}({\bm k},\eta )e^{-i\varphi}
			\right)
	\,,\label{eq:spin-0 sigma_g}\\
	&_{\pm 1}\sigma_\rmg
		\equiv\sigma_{\rmg ,i}e^i_\pm
		=\int\frac{\dd^3{\bm k}}{(2\pi )^3}e^{-ik\chi\mu}
			\left(
				\sigma_\rmg^{(+1)}({\bm k},\eta )\left(\mu \mp 1\right) e^{i\varphi}
				+\sigma_\rmg^{(-1)}({\bm k},\eta )\left(\mu \pm 1\right) e^{-i\varphi}
			\right)
	\,.\label{eq:spin-1 sigma_g}
} 
Since the quantities $_0\sigma_\rmg$ and $_{\pm 1}\sigma_\rmg$ transform as spin-$0$ and $\pm 1$ quantities,
acting with the spin-raising/lowering operators, eq.~\eqref{eq:pspin def}, 
on $_0\sigma_\rmg$ and $_{\pm 1}\sigma_\rmg$ leads to
\al{ 
	&\pspin^2\left( _0\sigma_\rmg\right)
		=\left( 1-\mu^2\right)
			\left( -\pd_\mu +\frac{i}{1-\mu^2}\pd_\varphi\right)^2
			\left(_0\sigma_\rmg\right)
	\notag\\
	&\quad\quad\quad\ 
		=\int\frac{\dd^3{\bm k}}{(2\pi )^3}
			\biggl\{
				\left(\pd_\mu^2 e^{-ik\chi\mu}\right)\left( 1-\mu^2\right)^{3/2}
				\left(
					\sigma_\rmg^{(+1)}e^{i\varphi}+\sigma_\rmg^{(-1)}e^{-i\varphi}
				\right)
	\notag\\
	&\quad\quad\quad\quad\quad
			-2\left(\pd_\mu e^{-ik\chi\mu}\right)\sqrt{1-\mu^2}
				\left(
					\sigma_\rmg^{(+1)}\left(\mu -1\right) e^{i\varphi}
					+\sigma_\rmg^{(-1)}\left(\mu +1\right) e^{-i\varphi}
				\right)
			\biggr\}
	\,,\label{eq:pspin^2 _0sigma_g}\\
	&\mspin^2\left( _0\sigma_\rmg\right)
		=\left( 1-\mu^2\right)
			\left( -\pd_\mu -\frac{i}{1-\mu^2}\pd_\varphi\right)^2
			\left(_0\sigma_\rmg\right)
	\notag\\
	&\quad\quad\quad\ 
		=\int\frac{\dd^3{\bm k}}{(2\pi )^3}
			\biggl\{
				\left(\pd_\mu^2 e^{-ik\chi\mu}\right)\left( 1-\mu^2\right)^{3/2}
				\left(
					\sigma_\rmg^{(+1)}e^{i\varphi}+\sigma_\rmg^{(-1)}e^{-i\varphi}
				\right)
	\notag\\
	&\quad\quad\quad\quad\quad
			-2\left(\pd_\mu e^{-ik\chi\mu}\right)\sqrt{1-\mu^2}
				\left(
					\sigma_\rmg^{(+1)}\left(\mu +1\right) e^{i\varphi}
					+\sigma_\rmg^{(-1)}\left(\mu -1\right) e^{-i\varphi}
				\right)
			\biggr\}
	\,,\label{eq:mspin^2 _0sigma_g}
}
and
\al{
	&\pspin\left(_{+1}\sigma_\rmg\right)
		=-\left( 1-\mu^2\right)\left( -\pd_\mu +\frac{i}{1-\mu^2}\pd_\varphi\right)
			\frac{1}{\sqrt{1-\mu^2}}\left(_{+1}\sigma_\rmg\right)
	\notag\\
	&\quad\quad\quad\ 
		=\int\frac{\dd^3{\bm k}}{(2\pi )^3}\left(\pd_\mu e^{-ik\chi\mu}\right)\sqrt{1-\mu^2}
			\left(
				\sigma_\rmg^{(+1)}\left(\mu -1\right) e^{i\varphi}
				+\sigma_\rmg^{(-1)}\left(\mu +1\right) e^{-i\varphi}
			\right)
	\,,\label{eq:pspin _+1sigma_g}\\
	&\mspin\left(_{-1}\sigma_\rmg\right)
		=-\left( 1-\mu^2\right)\left( -\pd_\mu -\frac{i}{1-\mu^2}\pd_\varphi\right)
			\frac{1}{\sqrt{1-\mu^2}}\left(_{-1}\sigma_\rmg\right)
	\notag\\
	&\quad\quad\quad\ 
		=\int\frac{\dd^3{\bm k}}{(2\pi )^3}\left(\pd_\mu e^{-ik\chi\mu}\right)\sqrt{1-\mu^2}
			\left(
				\sigma_\rmg^{(+1)}\left(\mu +1\right) e^{i\varphi}
				+\sigma_\rmg^{(-1)}\left(\mu -1\right) e^{-i\varphi}
			\right)
	\,,\label{eq:mspin _-1sigma_g}
} 
where $\pd_\mu\equiv\pd /\pd\mu$\,.
Plugging eqs.~\eqref{eq:pspin^2 _0sigma_g} and \eqref{eq:pspin _+1sigma_g} into
eq.~\eqref{eq:reduced shear in spin op}, we obtain the reduced shear
induced by vector perturbations:
\al{ 
	g=&-\frac{1}{2}\3Dint{\bm k}\int^{\cS}_0\frac{\dd\chi}{\chi}\sqrt{1-\mu^2}
			\Biggl[
				\frac{\cS -\chi}{\cS}\left( 1-\mu^2\right)
					\left(\pd_\mu^2\, e^{-ik\chi\mu}\right)
					\left(
						\sigma^{(+1)}_\rmg e^{i\varphi}+\sigma^{(-1)}_\rmg e^{-i\varphi}
					\right)
	\notag\\
			&+\left( 1-2\frac{\cS -\chi}{\cS}\right)
				\left(\pd_\mu\, e^{-ik\chi\mu}\right)
				\biggl\{
					\sigma^{(+1)}_\rmg\left(\mu -1\right) e^{i\varphi}
						+\sigma^{(-1)}_\rmg\left(\mu +1\right) e^{-i\varphi}
				\biggr\}
			\Biggr]
	\,.\label{eq:g}
} 
Following the same step as derived in the case of the spin-$+2$ part, 
we multiply $e^a_-e^b_-$ in both side of eq.~\eqref{eq:linear order Jacobi map solution} 
to construct the spin-$-2$ part of the reduced shear eq.~\eqref{eq:reduced shear def}:
\al{
	g^*=-\frac{1}{2}\int^{\cS}_0\frac{\dd\chi}{\chi}
		\biggl\{
			\frac{\cS -\chi}{\cS}\mspin^2\left(_0\sigma_\rmg\right)
			+\mspin\left(_{-1}\sigma_\rmg\right)
		\biggr\}
	\,,\label{eq:reduced shear2 in spin op}
}
where we have used eqs.~\eqref{eq:spin op2}, \eqref{eq:spin op4}.
Plugging eqs.~\eqref{eq:mspin^2 _0sigma_g} and \eqref{eq:mspin _-1sigma_g}
into eq.~\eqref{eq:reduced shear2 in spin op}, we have
\al{
	g^*=&-\frac{1}{2}\3Dint{\bm k}\int^{\cS}_0\frac{\dd\chi}{\chi}\sqrt{1-\mu^2}
		\Biggl[
			\frac{\cS -\chi}{\cS}\left( 1-\mu^{2}\right)
			\left(\pd_{\mu}^2\,e^{-ik\chi\mu}\right)
			\left(
				\sigma^{(+1)}_\rmg e^{i\varphi}+\sigma^{(-1)}_\rmg e^{-i\varphi}
			\right)
	\notag\\
		&+\left( 1-2\frac{\cS -\chi}{\cS}\right)
			\left(\pd_\mu\, e^{-ik\chi\mu}\right)
			\biggl\{
				\sigma^{(+1)}_\rmg\left(\mu +1\right) e^{i\varphi}
					+\sigma^{(-1)}_\rmg\left(\mu -1\right) e^{-i\varphi}
			\biggr\}
		\Biggr]
	\,.\label{eq:g^*}
}
We use the spin-$0$ quantities, $\tilde g^\rmE$ and $\tilde g^\rmB$ defined 
in eqs.~\eqref{eq:tilde g^E def} and \eqref{eq:tilde g^B def}\,,
to compute the angular power spectrum.
Since $g$ and $g^*$ transform as spin-$\pm 2$ quantities, we can 
apply the formula eqs.~\eqref{eq:mspin^2 spin+2X}, \eqref{eq:pspin^2 spin-2X}
to eqs.~\eqref{eq:g} and \eqref{eq:g^*} and calculate $\tilde g^\rmE$ and $\tilde g^\rmB$ as 
\al{ 
	\tilde g^\rmE
		=&\frac{1}{2}\3Dint{\bm k}\int^{\cS}_0\frac{\dd\chi}{\chi}\sqrt{1-\mu^2}
			\left(
				\sigma^{(+1)}_\rmg e^{i\varphi}+\sigma^{(-1)}_\rmg e^{-i\varphi}
			\right)
	\nonumber\\
	&\quad\quad\quad\quad\quad
	\times
		\left\{
			\frac{\cS -\chi}{\cS}\hat\mcE_0 (x)+\left( 1-2\frac{\cS -\chi}{\cS}\right)\hat\mcE_1 (x)
		\right\} e^{-ix\mu}\biggl|_{x=k\chi}
	\,,\label{eq:tilde g^E}\\
	\tilde g^\rmB
		=&-\frac{1}{2}\3Dint{\bm k}\int^{\cS}_0\frac{\dd\chi}{\chi}\sqrt{1-\mu^2}
			\left(
				\sigma^{(+1)}_\rmg e^{i\varphi}-\sigma^{(-1)}_\rmg e^{-i\varphi}
			\right)
			\hat\mcB (x)e^{-ix\mu}\biggl|_{x=k\chi}
	\,,\label{eq:tilde g^B}
} 
where we have introduced the operators $\hat\mcE_{0,1}(x)$ and $\hat\mcB (x)$ satisfying
the following equations:
\al{ 
	&\left( -\pd_\mu\pm\frac{1}{1-\mu^2}\right)^2
		\biggl[
			\left( 1-\mu^2\right)^{5/2}\pd_{\mu}^2\,e^{-ix\mu}
		\biggr]
	=\sqrt{1-\mu^2}
		\left(
			\hat\mcE_0 (x)\mp 2i\hat\mcB (x)
		\right) e^{-ix\mu}
	\,,\\
	&\left( -\pd_\mu\pm\frac{1}{1-\mu^2}\right)^2
		\biggl[
			\left( 1-\mu^2\right)^{3/2}\left(\mu\mp 1\right)\pd_\mu\,e^{-ix\mu}
		\biggr]
	=\sqrt{1-\mu^2}
		\left(
			\hat\mcE_1 (x)\mp i\hat\mcB (x)
		\right) e^{-ix\mu}\,.
} 
It follows that
\al{ 
	&\hat\mcE_0 (x) 
		=x^2\Bigl[
				4+20\pd_x^2 +10x\left(\pd_x +\pd_x^3\right) 
				+x^2\left( 1+\pd_x^2\right)^2
			\Bigr]
	\,,\\
	&\hat\mcE_1 (x) 
		=x\Bigl[
			-12\pd_x -x^2\left(\pd_x^3 +\pd_x\right) 
			-4x\left( 1+2\pd_x^2\right)
			\Bigr]
	\,,\\
	&\hat\mcB (x) 
		=x^2\Bigl[
			4\pd_x+x\left( 1+\pd_x^2\right)
			\Bigr]
	\,.
}
To obtain the multipole coefficients of $\tilde g^\rmE$ and $\tilde g^\rmB$\,,
we perform the angular integration using eqs.~\eqref{eq:angular integration of Y_lm}:
\al{
	\tilde g^\rmE_{\ell m}
		=&\int\dd^2\hatn\, Y_{\ell m}^*(\hatn )\,\tilde g^\rmE (\hatn )
	\notag\\
		=&(-i)^{\ell +1}
			\sqrt{4\pi (2\ell +1)\frac{(\ell +1)!}{(\ell -1)!}}
			\3Dint{\bm k}\int^{\cS}_0\frac{\dd\chi}{\chi}
				\left(
					\sigma_\rmg^{(+1)}\delta_{m,+1}-\sigma_\rmg^{(-1)}\delta_{m,-1}
				\right)
	\notag\\
		&\quad\quad
			\times
				\frac{1}{2}\left\{
					\frac{\cS -\chi}{\cS}\hat\mcE_0 (x)
					+\left( 1-2\frac{\cS -\chi}{\cS}\right)\hat\mcE_1 (x)
				\right\}
			\frac{j_\ell (x)}{x}\biggl|_{x=k\chi}
	\,,\label{eq:g^E 1}\\
	\tilde g^\rmB_{\ell m}
		=&\int\dd^2\hatn\, Y_{\ell m}^*(\hatn )\,\tilde g^\rmB (\hatn )
	\notag\\
		=&-(-i)^{\ell +1}
			\sqrt{4\pi (2\ell +1)\frac{(\ell +1)!}{(\ell -1)!}}
			\3Dint{\bm k}\int^{\cS}_0\frac{\dd\chi}{\chi}
	\notag\\
		&\quad\quad\quad 
			\times			
				\left(
					\sigma_\rmg^{(+1)}\delta_{m,+1}+\sigma_\rmg^{(-1)}\delta_{m,-1}
				\right)
				\frac{1}{2}\hat\mcB (x)\frac{j_\ell (x)}{x}\biggl|_{x=k\chi}
	\,.\label{eq:g^B 1}
}
To proceed, we act with $\hat\mcE_{0,1}(x)$ and $\hat\mcB (x)$ on $j_\ell (x)/x$:
\al{
	&\hat\mcE_0 (x)\,\frac{j_\ell (x)}{x}
		=\frac{(\ell -1)!(\ell +2)!}{(\ell +1)!(\ell -2)!}
			\left\{
				(\ell +2)(\ell -1)\frac{j_\ell (x)}{x}-2j_\ell' (x)
			\right\}
	\,,\\
	&\hat\mcE_1 (x)\,\frac{j_\ell (x)}{x}
		=-\frac{(\ell -1)!(\ell +2)!}{(\ell +1)!(\ell -2)!}
			\left\{
				\frac{j_\ell (x)}{x}+j_\ell' (x)
			\right\}
	\,,\\
	&\hat\mcB (x)\,\frac{j_\ell (x)}{x}
		=\frac{(\ell -1)!(\ell +2)!}{(\ell +1)!(\ell -2)!}\,j_\ell (x)
	\,,
}
where we have used the differential equation for the spherical Bessel function
(see Appendix \ref{sec:Spherical Bessel function}).
Since the multipole coefficients of the E-/B-mode cosmic shear are directly related
to those of $\tilde g^\rmE$ and $\tilde g^\rmB$ through eqs.~\eqref{eq:tilde g^E and g^E}
and \eqref{eq:tilde g^B and g^B}\,, we obtain
\al{
	g^\rmE_{\ell m}
		&=\sqrt{\frac{(\ell -2)!}{(\ell +2)!}}\,\tilde g^\rmE_{\ell m}
	\notag\\
	&=(-i)^{\ell +1}
			\sqrt{\pi (2\ell +1)\frac{(\ell -1)!(\ell +2)!}{(\ell +1)!(\ell -2)!}}
			\3Dint{\bm k}\int^{\cS}_0\frac{\dd\chi}{\chi}
				\left(
					\sigma_\rmg^{(+1)}\delta_{m,+1}-\sigma_\rmg^{(-1)}\delta_{m,-1}
				\right)
	\notag\\
	&\quad\times
			\biggl\{
				\frac{\cS -\chi}{\cS}\ell (\ell +1)\frac{j_\ell (x)}{x}
				-\left(\frac{j_\ell (x)}{x}+j_\ell' (x)\right)
			\biggr\}\biggl|_{x=k\chi}
	\notag\\
	&=(-i)^{\ell +1}
			\sqrt{4\pi (2\ell +1)}
			\3Dint{\bm k}\int^{\cS}_0 k\,\dd\chi
				\left(
					\sigma_\rmg^{(+1)}\delta_{m,+1}-\sigma_\rmg^{(-1)}\delta_{m,-1}
				\right)
				S^{\rm vector}_{\rmE ,\ell}(k,\chi )	
	\,,\label{eq:g^E_lm result}
}
and
\al{
	g^\rmB_{\ell m}
		&=\sqrt{\frac{(\ell -2)!}{(\ell +2)!}}\,\tilde g^\rmB_{\ell m}
	\notag\\
		&=-(-i)^{\ell +1}\sqrt{\pi (2\ell +1)\frac{(\ell -1)!(\ell +2)!}{(\ell +1)!(\ell -2)!}}
			\3Dint{\bm k}\int^{\cS}_0\frac{\dd\chi}{\chi} 
				\left(
					\sigma_\rmg^{(+1)}\delta_{m,+1}+\sigma_\rmg^{(-1)}\delta_{m,-1}
				\right)
				j_\ell (k\chi )
	\notag\\
		&=-(-i)^{\ell +1}\sqrt{4\pi (2\ell +1)}
			\3Dint{\bm k}\int^{\cS}_0 k\,\dd\chi 
				\left(
					\sigma_\rmg^{(+1)}\delta_{m,+1}+\sigma_\rmg^{(-1)}\delta_{m,-1}
				\right)
				S^{\rm vector}_{\rmB ,\ell}(k,\chi )
	\,.\label{eq:g^B_lm result}
}
where $S^{\rm vector}_{\rmE ,\ell}$, $S^{\rm vector}_{\rmB ,\ell}$ are defined
as
\al{
	&S_{\rmE ,\ell}^{\rm vector}(k,\chi )
		\equiv\frac{\cS -\chi}{\cS}\epsilon_\ell^{(0)}(k\chi )
			-\epsilon_\ell^{(1)}(k\chi )
	\,,\\
	&S_{\rmB ,\ell}^{\rm vector}(k,\chi )
		\equiv\beta_\ell^{(1)}(k\chi )
	\,,
}
with the coefficients $\epsilon_\ell^{(0,1)}$ and $\beta_\ell^{(1)}$ given by
\al{
	\epsilon^{(0)}_\ell (x)
		=&\frac{1}{2}\sqrt{\frac{(\ell -1)!}{(\ell +1)!}\frac{(\ell +2)!}{(\ell -2)!}}
			\,\ell (\ell +1)\frac{j_\ell (x)}{x^2}
	\,,\\
	\epsilon^{(1)}_\ell (x)
		=&\frac{1}{2}\sqrt{\frac{(\ell -1)!}{(\ell +1)!}\frac{(\ell +2)!}{(\ell -2)!}}
			\,\left(\frac{j_\ell (x)}{x^2}+\frac{j'_\ell (x)}{x}\right)
	\,,\\
	\beta^{(1)}_\ell (x)
		=&\frac{1}{2}\sqrt{\frac{(\ell -1)!}{(\ell +1)!}\frac{(\ell +2)!}{(\ell -2)!}}\,
			\frac{j_\ell (x)}{x}
	\,.
}
Substituting eqs.~\eqref{eq:g^E_lm result}, \eqref{eq:g^B_lm result}
into eq.~\eqref{eq:def_C_ell_EB},
we obtain the final expression for the angular power spectrum for 
the E-/B-mode cosmic shear:
\al{
	C_\ell^{\rm XX}
		=&\frac{2}{\pi}\int^\infty_0 k^2\,\dd k
			\int^{\cS}_0 k\,\dd\chi\int^{\cS}_0 k\,\dd\chi'
			S^{\rm vector}_{{\rm X},\ell}(k,\chi )
			S^{\rm vector}_{{\rm X},\ell}(k,\chi' )
			P_{\sigma_\rmg\sigma_\rmg}(k;\eta_0 -\chi ,\eta_0 -\chi' )
	\,,\\
	C_\ell^{\rm EB}=&0
	\,,
}
where we have used the condition for the un-polarized state of vector perturbations, 
eq.~\eqref{eq:power spectrum of vector perturbation def}.
If we consider the polarized state of vector perturbations, 
the nonzero cross correlation between E- and B-mode would appear.

\subsection{Scalar-/pseudo-scalar lensing potential} 
\label{sec:Scalar/pseudo-scalar lensing potential}

In this subsection, we derive the explicit expression for the scalar-/pseudo-scalar
lensing potentials and the gradient-/curl-mode angular power spectra
induced by vector perturbations.
The calculation of the angular power spectra are basically the same way as in the case
of cosmic shear in previous subsection.
In terms of the spin operators, eqs.~\eqref{eq:scalar lensing potential in spin op}
and \eqref{eq:pseudo-scalar lensing potential in spin op} are
\al{
	\nabla^2\phi
		=&\int^{\cS}_0\frac{\dd\chi}{\chi}
			\biggl[
				\frac{\cS -\chi}{\cS}\,\pspin\mspin\left(_0\sigma_\rmg\right)
				+\frac{1}{2}
					\biggl\{
						\mspin\,\left(_{+1}\sigma_\rmg\right)
						+\pspin\,\left(_{-1}\sigma_\rmg\right)
					\biggr\}
			\biggr]
	\,,\label{eq:nabla^2 phi}\\
	\nabla^2\varpi 
		=&\,\frac{1}{2i}\int^{\cS}_0\frac{\dd\chi}{\chi}
			\Bigl\{
				\mspin\left(_{+1}\sigma_\rmg\right) -\pspin\left(_{-1}\sigma_\rmg\right)
			\Bigr\}
	\,,\label{eq:nabla^2 varpi}
}
where $_0\sigma_\rmg =\sigma_{\rmg ,i}e^i_\chi$\,, $_{\pm 1}\sigma_\rmg =\sigma_{\rmg ,i}e^i_\pm$\,,
and we have used eqs.~\eqref{eq:spin op3}, \eqref{eq:spin op5}\,.
In the coordinate $k^i\equiv k\hat k^i=k(0,0,1)$ and $e^i_\pm (\hatk )=(1,\pm i,0)$\,,
we act with the spin-raising/lowering operators on 
$_0\sigma_\rmg$ and $_{\pm 1}\sigma_\rmg$ (eqs.~\eqref{eq:spin-0 sigma_g}, 
and \eqref{eq:spin-1 sigma_g}):
\al{
	&\pspin\mspin\left(_0\sigma_\rmg\right)
		=\int\frac{\dd^3{\bm k}}{(2\pi )^3}
			\sqrt{1-\mu^2}
			\left(\sigma_\rmg^{(+1)}e^{i\varphi}+\sigma_\rmg^{(-1)}e^{-i\varphi}\right)
			\Bigl( (1-\mu^2 )\pd_\mu^2 -4\mu\,\pd_\mu -2\Bigr) e^{-ik\chi\mu}
	\,,\label{eq:pspin mspin _0sigma_g}\\
	&\mspin\left(_{+1}\sigma_\rmg\right)
		=\int\frac{\dd^3{\bm k}}{(2\pi )^3}\sqrt{1-\mu^2}
			\biggl[
				\left(\pd_\mu e^{-ik\chi\mu}\right)
				\left(\sigma_\rmg^{(+1)}(\mu -1)e^{i\varphi}+\sigma_\rmg^{(-1)}(\mu +1)e^{-i\varphi}\right)
	\notag\\
	&\quad\quad\quad\quad\quad\quad\quad
				+2e^{-ik\chi\mu}\left(\sigma_\rmg^{(+1)}e^{i\varphi}+\sigma_\rmg^{(-1)}e^{-i\varphi}\right)
			\biggr]
	\,,\label{eq:mspin _+1sigma_g}\\
	&\pspin\left(_{-1}\sigma_\rmg\right)
		=\int\frac{\dd^3{\bm k}}{(2\pi )^3}\sqrt{1-\mu^2}
			\biggl[
				\left(\pd_\mu e^{-ik\chi\mu}\right)
				\left(\sigma_\rmg^{(+1)}(\mu +1)e^{i\varphi}+\sigma_\rmg^{(-1)}(\mu -1)e^{-i\varphi}\right)
	\notag\\
	&\quad\quad\quad\quad\quad\quad\quad
				+2e^{-ik\chi\mu}\left(\sigma_\rmg^{(+1)}e^{i\varphi}+\sigma_\rmg^{(-1)}e^{-i\varphi}\right)
			\biggr]
	\,.\label{eq:pspin _-1sigma_g}
}
Plugging eqs.~\eqref{eq:pspin mspin _0sigma_g}-\eqref{eq:pspin _-1sigma_g} 
into eqs.~\eqref{eq:nabla^2 phi}, \eqref{eq:nabla^2 varpi},
we rewrite the scalar/pseudo-scalar lensing potentials with
\al{
	\nabla^2\phi
		=&\3Dint{\bm k}\int^{\cS}_0\frac{\dd\chi}{\chi}\sqrt{1-\mu^2}
			\left(\sigma_\sigma^{(+1)}e^{i\varphi}+\sigma_\rmg^{(-1)}e^{-i\varphi}\right)
	\notag\\
			&\quad\quad\times
				\biggl\{
					\frac{\cS -\chi}{\cS}\Bigl( (1-\mu^2 )\pd_\mu^2 -4\mu\pd_\mu -2\Bigr) 
					+\Bigl( \mu\pd_\mu +2\Bigr)
				\biggr\} e^{-ix\mu}\biggl|_{x=k\chi}
	\,,\\
	\nabla^2\varpi 
		=&\,i\3Dint{\bm k}\int^{\cS}_0\frac{\dd\chi}{\chi}
			\sqrt{1-\mu^2}\left(\pd_\mu e^{-ik\chi\mu}\right)
			\left(
				\sigma^{(+1)}_\rmg e^{i\varphi}-\sigma^{(-1)}_\rmg e^{-i\varphi}
			\right)
	\notag\\
		=&\3Dint{\bm k}\int^{\cS}_0k\,\dd\chi
			\sqrt{1-\mu^2}
			\left(
				\sigma^{(+1)}_\rmg e^{i\varphi}-\sigma^{(-1)}_\rmg e^{-i\varphi}
			\right) 
			e^{-ik\chi\mu}
	\,.
}
Performing the angular integration and using the differential equation
for the spherical Bessel function (see Appendix \ref{sec:Spherical Bessel function}),
the multipole coefficients of $\phi$ and $\varpi$ are expressed as
\al{
	\phi_{\ell m}
		=&\int\dd^2\hatn\, Y_{\ell m}^*(\hatn )\,\phi (\hatn )
	\notag\\
	\quad
		=&(-i)^{\ell +1}\sqrt{4\pi (2\ell +1)\frac{(\ell -1)!}{(\ell +1)!}}
			\int\frac{\dd^3{\bm k}}{(2\pi )^3}\int^{\cS}_0\frac{\dd\chi}{\chi}
				\left(\sigma_\rmg^{(+1)}\delta_{m,+1}-\sigma_\rmg^{(-1)}\delta_{m,-1}\right)
	\notag\\
	&\quad\quad\quad\quad\times
			\biggl\{
					\frac{\cS -\chi}{\cS}\Bigl( x^2\left( 1+\pd_x^2\right) +4x\pd_x +2\Bigr) 
					-\Bigl( x\pd_x +2\Bigr)
			\biggr\}
			\frac{j_\ell (x)}{x}\biggl|_{x=k\chi}
	\notag\\
		=&(-i)^{\ell +1}\sqrt{4\pi (2\ell +1)\frac{(\ell -1)!}{(\ell +1)!}}
			\int\frac{\dd^3{\bm k}}{(2\pi )^3}\int^{\cS}_0\frac{\dd\chi}{\chi}
				\left(\sigma_\rmg^{(+1)}\delta_{m,+1}-\sigma_\rmg^{(-1)}\delta_{m,-1}\right)
	\notag\\
	&\quad\quad\quad\quad\times
			\biggl\{
					\frac{\cS -\chi}{\cS}\ell (\ell +1)\frac{j_\ell (x)}{x}
					-\left(\frac{j_\ell (x)}{x}+j_\ell' (x)\right)
			\biggr\}\biggl|_{x=k\chi}
	\notag\\
		=&(-i)^{\ell +1}
			\sqrt{4\pi (2\ell +1)}
			\3Dint{\bm k}\int^{\cS}_0\, k\,\dd\chi
				\left(
					\sigma_\rmg^{(+1)}\delta_{m,+1}-\sigma_\rmg^{(-1)}\delta_{m,-1}
				\right)
				S^{\rm vector}_{\phi ,\ell}(k,\chi )
	\,,\label{eq:phi_ell_m_sigma}
}
and
\al{
	\varpi_{\ell m}
		=&\int\dd^2\hatn\, Y_{\ell m}^*(\hatn )\,\varpi (\hatn )
	\notag\\
		=&-(-i)^{\ell +1}\sqrt{4\pi (2\ell +1)\frac{(\ell -1)!}{(\ell +1)!}}
			\int\frac{\dd^3{\bm k}}{(2\pi )^3}\int^{\cS}_0\frac{\dd\chi}{\chi}
				\left(\sigma_\rmg^{(+1)}\delta_{m,+1}+\sigma_\rmg^{(-1)}\delta_{m,-1}\right)\, j_\ell (k\chi )
	\notag\\
		=&-(-i)^{\ell +1}
			\sqrt{4\pi (2\ell +1)}
			\3Dint{\bm k}\int^{\cS}_0\, k\,\dd\chi 
				\left(
					\sigma_\rmg^{(+1)}\delta_{m,+1}+\sigma_\rmg^{(-1)}\delta_{m,-1}
				\right)
				S^{\rm vector}_{\varpi ,\ell}(k,\chi )
	\,,\label{eq:varpi_ell_m_sigma}
}
where we have used eq.~\eqref{eq:angular integration of Y_lm} and 
the fact $\nabla^2 Y_{\ell m}=-\ell (\ell +1)Y_{\ell m}$\,.
$S^{\rm vector}_{\phi ,\ell}$ and $S^{\rm vector}_{\varpi ,\ell}$ are defined by
\al{
	&S^{\rm vector}_{\phi ,\ell}(k,\chi )
		\equiv 2\,\sqrt{\frac{(\ell -2)!}{(\ell +2)!}}\,
		 \left(
			\frac{\cS -\chi}{\cS}\,\epsilon^{(0)}_\ell (k\chi )
			-\epsilon^{(1)}_\ell (k\chi )
		 \right)
	\,,\\
	&S^{\rm vector}_{\varpi ,\ell}(k,\chi )
		\equiv 2\,\sqrt{\frac{(\ell -2)!}{(\ell +2)!}}\,\beta^{(1)}_\ell (k\chi )
	\,,
}
with the quantities $\epsilon_\ell^{(0,1)}$ and $\beta_\ell^{(1)}$ 
given by eqs.~\eqref{eq:epsilon^0 def}-\eqref{eq:beta^1 def}. 
Substituting eqs.~\eqref{eq:phi_ell_m_sigma}, \eqref{eq:varpi_ell_m_sigma} into 
eq.~\eqref{eq:def_C_ell_phi}\,, and
using the condition for the un-polarized state of vector perturbations 
eq.~\eqref{eq:power spectrum of vector perturbation def}\,,
we obtain the explicit expression for the angular power spectra for 
the gradient-/curl-mode deflection angle:
\al{
	C_\ell^{xx}
		=&\frac{2}{\pi}\int^\infty_0 k^2\,\dd k
			\int^{\cS}_0 k\,\dd\chi\int^{\cS}_0 k\,\dd\chi'
			S^{\rm vector}_{x,\ell}(k,\chi )
			S^{\rm vector}_{x,\ell}(k,\chi' )
			P_{\sigma_\rmg\sigma_\rmg}(k;\eta_0 -\chi ,\eta_0 -\chi' )
	\,,\\
	C_\ell^{\phi\varpi}
		=&0
	\,.
}

\subsection{Derivation of shear-deflection relation} 
\label{sec:Derivation of shear-deflection relation}

In this subsection, we derive the explicit relation between cosmic shear
and deflection angle in terms of the spin operators.
The relation eq.~\eqref{eq:shear-lensing potential relations} 
can be further reduced to simplified forms if we move to the harmonic space. 
With a help of $\nabla^2 Y_{\ell m}(\hatn )=-\ell (\ell +1)Y_{\ell m}(\hatn )$, 
we have
\al{
	\phi_{\ell m}
		=&2\frac{(\ell -2)!}{(\ell +2)!}
			\int\dd^2\hatn\, Y_{\ell m:ab}^*\gamma_{cd}\omega^{ac}\omega^{bd}
	\,,\\
	\varpi_{\ell m}
		=&2\frac{(\ell -2)!}{(\ell +2)!}
			\int\dd^2\hatn\, Y_{\ell m:ab}^*\gamma_{cd}\omega^{ac}\epsilon^{bd}
	\,.
}
Using eqs.~\eqref{eq:shear def}\,, \eqref{eq:reduced shear def}\,, and \eqref{eq:spin op4}\,,
we then rewrite the metric on the sphere, $\omega^{ab}$\,, the Levi-Civita pseudo-tensor, $\epsilon^{ab}$\,,
in terms of the basis vector $e^a_\pm$ (see eq.~\eqref{eq:omega-epsilon-spin basis relations})
and spin operators:
\al{
	\phi_{\ell m}
		=&2\frac{(\ell -2)!}{(\ell +2)!}
			\int\dd^2\hatn\, Y_{\ell m:ab}^*\gamma_{cd}e^{(a}_+e^{c)}_-e^{(b}_+e^{d)}_-
	\notag\\
		=&-\frac{(\ell -2)!}{(\ell +2)!}
			\int\dd^2\hatn\,
				\biggl[
					\left( \pspin^2 Y_{\ell m}(\hatn )\right)^* g(\hatn )
					+\left( \mspin^2 Y_{\ell m}(\hatn )\right)^* g^*(\hatn )
				\biggr]
	\,,\\
	\varpi_{\ell m}
		=&2i\frac{(\ell -2)!}{(\ell +2)!}
			\int\dd^2\hatn\, Y_{\ell m:ab}^*\gamma_{cd}e^{(a}_+e^{c)}_-e^{[b}_+e^{d]}_-
		\notag\\
		=&i\frac{(\ell -2)!}{(\ell +2)!}
			\int\dd^2\hatn\,
				\biggl[
					\left( \pspin^2 Y_{\ell m}(\hatn )\right)^* g(\hatn )
					-\left( \mspin^2 Y_{\ell m}(\hatn )\right)^* g^*(\hatn )
				\biggr]
	\,,
}
where we have used the traceless condition for the shear, 
namely $\gamma_{ab}\omega^{ab}=\gamma_{ab}e^a_+e^b_-=0$\,.
We then rewrite $\pspin^2 Y_{\ell m}$ and $\mspin^2 Y_{\ell m}$
in terms of the spin-$\pm 2$ spherical harmonics ${}_{\pm 2}Y_{\ell m}$
(see eqs.~\eqref{eq:spin-weighted positive spherical harmonics}):
\al{
	\phi_{\ell m}
		=&-\sqrt{\frac{(\ell -2)!}{(\ell +2)!}}
			\int\dd^2\hatn\,
				\Bigl\{
					{}_{+2}Y_{\ell m}^*(\hatn )\, g(\hatn )
					+{}_{-2}Y_{\ell m}^*(\hatn )\, g^*(\hatn )
				\Bigr\}
	\notag\\
		=&-\sqrt{\frac{(\ell -2)!}{(\ell +2)!}}
			\Bigl({}_{+2}g_{\ell m}+{}_{-2}g_{\ell m}\Bigr)
	\,,\\
	\varpi_{\ell m}
		=&i\sqrt{\frac{(\ell -2)!}{(\ell +2)!}}
			\int\dd^2\hatn\,
				\Bigl\{
					{}_{+2}Y_{\ell m}^*(\hatn )\, g(\hatn )
					-{}_{-2}Y_{\ell m}^*(\hatn )\, g^*(\hatn )
				\Bigr\}
	\notag\\
		=&i\sqrt{\frac{(\ell -2)!}{(\ell +2)!}}
			\Bigl({}_{+2}g_{\ell m}-{}_{-2}g_{\ell m}\Bigr)
	\,.
} 
Recalling that the combination $({}_{+2}g_{\ell m}\pm{}_{-2}g_{\ell m})$ can be
rewritten in terms of the E-/B-mode cosmic shear field (see eq.~\eqref{eq:E-/B-mode def})\,, 
we obtain the explicit relations between $\phi_{\ell m}$\,, $\varpi_{\ell m}$\,, $g^\rmE_{\ell m}$\,,
and $g^\rmB_{\ell m}$ (see also \cite{Stebbins:1996wx}):
\al{
	\phi_{\ell m}
		=2\sqrt{\frac{(\ell -2)!}{(\ell +2)!}}\,g_{\ell m}^\rmE
	\,,\ \ \ 
	\varpi_{\ell m}
		=2\sqrt{\frac{(\ell -2)!}{(\ell +2)!}}\,g_{\ell m}^\rmB
	\,.
} 

\section{Derivation of correlations of a cosmic string network} 
\label{sec:Derivation of correlations of a cosmic string network}

Let us consider a Nambu-Goto string segment at the position ${\bm r} ={\bm r} (\sigma ,\eta )$
where $\eta$ and $\sigma$ are the time and position on the string worldsheet.
In the transverse gauge, the stress-energy tensor for a string segment can be described 
as~\cite{Vilenkin-Shellard}
\beq
	\delta T^{\mu\nu}({\bm r},\eta )
		=\mu\int\dd\sigma
			\left(
			\begin{array}{cc}
				1 & -\dot r^i\\
				-\dot r^j & \dot r^i\dot r^j -{r^i}'{r^j}'\\
			\end{array}
			\right)
			\delta^{3} ({\bm r}-{\bm r}(\sigma ,\eta ))
	\,,\label{eq:string stress-energy}
\eeq
where the dot ( $\dot{}$ ) and the prime ( ${}'$ ) denote the derivative with respect to $\eta$ and $\sigma$.
Comparing to eqs.~\eqref{eq:velocity perturbations} and \eqref{eq:string stress-energy},
the velocity perturbations, $v^{(\pm 1)}$, due to a segment are given by
\beq
	v^{(\pm 1)}({\bm k},\eta )
		=\mu\int\dd\sigma\,\dot r^i (\sigma ,\eta )\,e_{\pm ,i}^* (\hatk )\,
			e^{i{\bm k}\cdot{\bm r} (\sigma ,\eta )}
	\,.\label{eq:string velocity perturbations}
\eeq
Since the correlations can be described by 
a summation of the contribution of each segment,
we can estimate the equal-time auto-power spectrum for the vector perturbations as
\al{
	P_{\sigma_\rmg\sigma_\rmg}(k;\eta ,\eta )
		=&2\frac{(16\pi G)^2a^4}{k^4}
			\frac{1}{\mcV}
				\ave{
					v^{(\pm 1)*}({\bm k},\eta )v^{(\pm 1)}({\bm k},\eta )
				}
	\notag\\
		=&2\frac{(16\pi G\mu )^2a^4}{k^4}
				n_{\rm s}\,\dd V\,
				\frac{1}{\mcV}\,e_{\pm ,i}^*(\hatk )e_{\pm ,j}(\hatk )
	\notag\\
	&\quad\quad\quad\times
				\ave{
					\int\dd\sigma_1\dd\sigma_2\,
					\dot r^i(\sigma_1 ,\eta )\,\dot r^j(\sigma_2 ,\eta )\,
					e^{i{\bm k}\cdot ({\bm r}(\sigma_1 ,\eta )-{\bm r}(\sigma_2 ,\eta ))}
				}\,.
}
where $\dd V =4\pi\chi^2/H$ is the differential comoving volume element, 
$n_{\rm s}=a^3\xi^{-3}$ is the comoving number density of string segments, 
and $\mcV =(2\pi )^3\delta^3 ({\bm 0})$ is the comoving box size.
For the string averaging, we can use a very simple model
developed in \cite{Vincent:1996qr,Albrecht:1997mz,Hindmarsh:1993pu}.
The assumption in this model is that all correlators can be expressed in terms of
two-point correlations for $\dot r^i(\sigma ,\eta )$ and ${r^i}'(\sigma ,\eta )$\,.
Assuming that $\dot r^i (\sigma ,\eta )$ and ${r^i}' (\sigma ,\eta )$ are exactly 
Gaussian and isotropic distributed with mean zero and variances 
$\ave{\dot r^i(\sigma ,\eta)\dot r^j(0,\eta )}\equiv\frac{1}{3}\delta^{ij}\,V_{\rm s}(\sigma )$, 
and $\big\langle{r^i}'(\sigma ,\eta ){r^j}'(0,\eta )\big\rangle\equiv\frac{1}{3}\delta^{ij}\,T_{\rm s}(\sigma )$\,,
we can compute the equal-time auto-power spectrum for the vector perturbations as
\beq
	P_{\sigma_\rmg\sigma_\rmg}(k,\eta ,\eta )
		=\frac{(16\pi G\mu )^2a^4}{k^4}
			n_{\rm s}\,\dd V\,
			\frac{1}{3\mcV}
			\int\dd\sigma_+\dd\sigma_-
				V_{\rm s}(\sigma_- )
				\exp\biggl[-\frac{1}{6}k^2\Gamma_{\rm s} (\sigma_- )\biggr]
	\,,
	\label{eq:averaged power spectrum}
\eeq
where $\sigma_\pm =\sigma_1\pm\sigma_2$, we have introduced
$\Gamma_{\rm s} (\sigma )=\int^\sigma_0\dd\sigma_3\dd\sigma_4 T_{\rm s}(\sigma_3 -\sigma_4 )$\,.
On scale larger than the correlation length, the correlators
are expected to be damped, and the correlators on scale $\sigma <\xi /a$
can be approximated as $V_{\rm s}\approx v_{\rm rms}^2$,
$\Gamma_{\rm s}\approx (1-v_{\rm rms}^2)\,\sigma^2$\,.
Once we determine the region of the integration, we can calculate 
the auto-power spectrum of the velocity perturbations 
by integrating eq.~\eqref{eq:averaged power spectrum}.
Since the term $\int\dd\sigma_+ /\mcV$ corresponds to the length 
of the string segment within the unit volume
and the correlators, $V_{\rm s}$ and $\Gamma_{\rm s}$, 
are damped at $\sigma_-\gg\xi /a$\,, 
we take the region of the integration as $\int\dd\sigma_+ /\mcV =a^2/\xi^2\sqrt{1-v_{\rm rms}^2}$
and $|\sigma_-|\leq \xi /2a\sqrt{1-v_{\rm rms}^2}$ hereafter. 
Then, we have
\beq
	P_{\sigma_\rmg\sigma_\rmg}(k;\eta ,\eta )
		\approx\,\left( 16\pi G\mu\right)^2
					\frac{2\sqrt{6\pi}\, v_{\rm rms}^2}{3(1-v_{\rm rms}^2)}
					\frac{4\pi\chi^2 a^4}{H}\left(\frac{a}{k\xi}\right)^5
					{\rm erf}\left(\frac{k\xi /a}{2\sqrt{6}}\right)
	\,.
\eeq

\section{Reconstruction noise}
\label{sec:Reconstruction noise}

\bc
\begin{table}[tbp]
\bc
\begin{tabular}{c|c|c} \hline 
${\rm XY}$ & $f^{\phi,(\alpha )}_{\ell,L,L'}$ & $f^{\varpi, (\alpha )}_{\ell,L,L'}$ 
\\ \hline \\
$\Theta\Theta$ & 
$_0\mcS^{\phi}_{L,\ell,L'}C_{L'}^{\Theta\Theta}+_0\mcS^{\phi}_{L',\ell,L}C_L^{\Theta\Theta}$ &
$_0\mcS^{\varpi}_{L,\ell,L'}C_{L'}^{\Theta\Theta}-_0\mcS^{\varpi}_{L',\ell,L}C_L^{\Theta\Theta}$ 
\\ \\ \hline \\ 
$\Theta\rmE$ & 
$_0\mcS^{\phi}_{L,\ell,L'}C_{L'}^{\Theta\rmE}+_{\oplus}\mcS^{\phi}_{L',\ell,L}C_L^{\Theta\rmE}$&
$_0\mcS^{\varpi}_{L,\ell,L'}C_{L'}^{\Theta\rmE}-_{\oplus}\mcS^{\varpi}_{L',\ell,L}C_L^{\Theta\rmE}$
\\ \\ \hline \\ 
$\Theta\rmB$ & 
$-_{\ominus}\mcS^{\phi}_{L',\ell,L}C_L^{\Theta\rmE}$&
$_{\ominus}\mcS^{\varpi}_{L',\ell,L}C_L^{\Theta\rmE}$ 
\\ \\ \hline \\
$\rmE\rmE$ & 
$_{\oplus}\mcS^{\phi}_{L,\ell,L'}C_{L'}^{\rmE\rmE}+_{\oplus}\mcS^{\phi}_{L',\ell,L}C_L^{\rmE\rmE}$&
$_{\oplus}\mcS^{\varpi}_{L,\ell,L'}C_{L'}^{\rmE\rmE}-_{\oplus}\mcS^{\varpi}_{L',\ell,L}C_L^{\rmE\rmE}$ 
\\ \\ \hline \\ 
$\rmE\rmB$ & 
$-_{\ominus}\mcS^{\phi}_{L,\ell,L'}C_{L'}^{\rmB\rmB}-_{\ominus}\mcS^{\phi}_{L',\ell,L}C_L^{\rmE\rmE}$& 
$-_{\ominus}\mcS^{\varpi}_{L,\ell,L'}C_{L'}^{\rmB\rmB}+_{\ominus}\mcS^{\varpi}_{L',\ell,L}C_L^{\rmE\rmE}$ 
\\ \\ \hline \\
$\rmB\rmB$ & 
$_{\oplus}\mcS^{\phi}_{L,\ell,L'}C_{L'}^{\rmB\rmB}+_{\oplus}\mcS^{\phi}_{L',\ell,L}C_L^{\rmB\rmB}$&
$_{\oplus}\mcS^{\varpi}_{L,\ell,L'}C_{L'}^{\rmB\rmB}-_{\oplus}\mcS^{\varpi}_{L',\ell,L}C_L^{\rmB\rmB}$ 
\\ \\ \hline 
\end{tabular}
\caption{The functional forms of $f^{x,(\alpha )}_{\ell,L,L'}$\,. }
\vs{0.5}
\label{table:f}
\ec
\end{table} 
\ec

\bc
\begin{table}[tbp]
\caption{
Experimental specifications for the PLANCK and ACTPol used in this paper. 
The quantity $\theta_\nu$ is the beam size, and $\sigma_{\nu}$ 
represents the sensitivity of each channel to the temperature 
$\sigma_{\nu,T}$ or polarizations $\sigma_{\nu,P}$. 
The quantity $\nu$ means a channel frequency. }
\label{CMB survey design}
\vs{0.5}
\bc
\begin{tabular}{cccccc} \hline 
Experiment & $f\rom{sky}$ & $\nu$ [GHz] & $\theta_{\nu}$ [arcmin] 
& $\sigma_{\nu,T}$ [$\mu$K/pixel] & $\sigma_{\nu,P}$ [$\mu$K/pixel] 
\\ \hline 
PLANCK~\cite{:2006uk} & 0.65 & 30 & 33 & 4.4 & 6.2 \\ 
 & & 44 & 23 & 6.5 & 9.2  \\ 
 & & 70 & 14 & 9.8 & 13.9 \\
 & & 100 & 9.5 & 6.8 & 10.9 \\
 & & 143 & 7.1 & 6.0 & 11.4 \\
 & & 217 & 5.0 & 13.1 & 26.7 \\
 & & 353 & 5.0 & 40.1 & 81.2 \\ \hline 
ACTPol~\cite{arXiv:1006.5049} & 0.1 & 148 & 1.4 & 3.6 & 5.0 \\ \hline 
\end{tabular}
\ec
\end{table} 
\ec

We provide the brief summary of the reconstruction noise spectrum, following \cite{Namikawa:2011cs}.
The reconstruction noise spectrum for the optimal combination of the minimum variance 
estimator is given by
\beq
	N_\ell^{x,(c)}
		=\Biggl[\,\sum_{\alpha ,\beta}\Bigl\{\left({\bm N}_\ell^x\right)^{-1}\Bigr\}_{\alpha\beta}\Biggr]^{-1}
	\,,\label{eq:optimal combination of minimum variance estimator}
\eeq
where $x=\phi ,\varpi$, the subscripts $\alpha ,\beta$ mean a pair of two CMB maps, and 
the component of the matrix $\{{\bm N}_\ell^x\}_{\alpha\beta}$ is the covariance
of the reconstruction noise, which is given by
\beq
	N_\ell^{x, (\alpha ,\beta )}
		=\frac{1}{2\ell +1}\sum_{L,L'}^{\ell_{\rm max}}\left( F_{\ell ,L,L'}^{x,(\alpha )}\right)^*
			\left( 
				F_{\ell ,L,L'}^{x,(\beta )}\tilde C_L^{\rm XX'}\tilde C_{L'}^{\rm YY'}
				+F_{\ell ,L,L'}^{x,(\beta )}(-1)^{\ell +L+L'}\tilde C_L^{\rm XY'}\tilde C_{L'}^{\rm X'Y}
			\right)
	\,,
\eeq
where $\ell_{\rm max}$ denotes the maximum multipole used in the reconstruction procedure.
The quantity $\tilde C_L^{\rm XY}$ is the lensed CMB angular power spectrum including
the contributions from instrumental noise and we have introduced 
the weight function $F_{\ell ,L,L'}^{x,(\alpha )}$ defined by
\beq
	F_{\ell ,L,L'}^{x,(\alpha )}=N_\ell^{x,(\alpha )}g_{\ell ,L,L'}^{x, (\alpha )}
	\,,
\eeq
where
\al{
	&N_\ell^{x,(\alpha )}
		=\Biggl[
			\frac{1}{2\ell +1}\sum_{L,L'}^{\ell_{\rm max}}\,f_{\ell ,L,L'}^{x,(\alpha )}\,g_{\ell ,L,L'}^{x,(\alpha )}
		 \Biggr]^{-1}
	\,,\\
	&g_{\ell, L,L'}^{x,(\alpha )}
		=\frac{\left( f_{\ell ,L,L'}^{x,(\alpha )}\right)^*\tilde C_{L'}^{\rm XX}\tilde C_L^{\rm YY}
			-(-1)^{\ell +L+L'}\left( f_{\ell ,L',L}^{x,(\alpha )}\right)^*\tilde C_L^{\rm XY}\tilde C_{L'}^{\rm XY}}
		{\tilde C_L^{\rm XX}\tilde C_{L'}^{\rm YY}\tilde C_{L'}^{\rm XX}\tilde C_L^{\rm YY}
			-\left(\tilde C_L^{\rm XY}\tilde C_{L'}^{\rm XY}\right)^2}
	\,.
}
The coefficients, $f_{\ell ,L,L'}^{x,(\alpha )}$\,, are expressed by the combination
of the unlensed CMB angular power spectrum, $C_\ell^{\rm XY}$\,, and 
the quantities, $_0\mcS^x_{\ell ,L,L'}$, $_\oplus\mcS^x_{\ell ,L,L'}$, and $_\ominus\mcS^x_{\ell ,L,L'}$.
We summarize $f_{\ell ,L,L'}^{x,(\alpha )}$ in Table \ref{table:f}.
The quantities, ${}_{0,\oplus,\ominus}\mcS^x_{\ell ,L,L'}$\,, are written 
in terms of the Wigner-$3j$ symbols as
\al{
	_s\mcS^\phi_{\ell ,L,L'}
		=&\sqrt{\frac{(2L +1)(2L'+1)(2\ell +1)}{16\pi}}
			\Bigl[L(L+1)+L'(L'+1)-\ell (\ell +1)\Bigr]
			\left(
				\begin{array}{ccc}
				\ell &L &L'\\
				s&0&-s\\
				\end{array}
			\right)
	\,,\\
	_s\mcS^\varpi_{\ell ,L,L'}
		=&\sqrt{\frac{(2L +1)(2L'+1)(2\ell +1)}{16\pi}}
			\sqrt{L (L +1)(L'+s)(L'+1-s)}
	\notag\\
		 &\times\Biggl[
			\sqrt{\frac{L'+1-s}{L'+1+s}}
				\left(
					\begin{array}{ccc}
					\ell &L &L'\\
					s&-1&1-s\\
					\end{array}
				\right)
			-\sqrt{\frac{L'-s}{L'+s}}
				\left(
					\begin{array}{ccc}
					\ell &L&L'\\
					s&1&-1-s\\
					\end{array}
				\right)
			\Biggr]
	\,,\\
	_\oplus\mcS^x_{\ell ,L,L'} 
		=&\frac{1}{2}\left(_2\mcS^x_{\ell ,L,L'}+_{-2}\mcS^x_{\ell ,L,L'}\right)
	\,,\ \ \ 
	_\ominus\mcS^x_{\ell ,L,L'}
		=\frac{1}{2i}\left(_2\mcS^x_{\ell ,L,L'}-_{-2}\mcS^x_{\ell ,L,L'}\right)
}
The instrumental noise is given by
\beq
	\mcN^{\rm XX}_\ell 
		=\biggl[\sum_\nu\left(\mcN^{XX}_{\ell ,\nu}\right)^{-1}\biggr]^{-1}
	\,;\ \ 
	\mcN^{\rm XX}_{\ell ,\nu}
		=\left(\frac{\sigma_\nu\theta_\nu}{T_{\rm CMB}}\right)^2
			\exp\biggl[
					\frac{\ell (\ell +1)\theta_\nu^2}{8\ln 2}
				\biggr]
	\,,
\eeq
where $T_{\rm CMB}=2.7{\rm K}$\,, $\theta_\nu$\,, and $\sigma_\nu$ represent
the mean temperature of CMB, the beam size, and the sensitivity of each channel.
We summarize the basic parameters for PLANCK and ACTPol in Table \ref{CMB survey design}.
For the cosmic variance limit, we take ${\mcN}_\ell^{\rm XX}=0$.
The reconstruction noise spectrum for ACTPol+PLANCK is assumed to have the form:
\beq
	N_{\ell :{\rm ACTPol+PLANCK}}^{\varpi ,(c)}
		=\left(
			\frac{f^{\rm ACTPol}_{\rm sky}}{\left( N^{\varpi ,(c)}_{\ell :{\rm ACTPol}}\right)^2}	
			+\frac{f^{\rm PLANCK}_{\rm sky}-f^{\rm ACTPol}_{\rm sky}}
				{\left( N^{\varpi ,(c)}_{\ell :{\rm PLANCK}}\right)^2}
		 \right)^{-1/2}
	\,,
\eeq
where $f^{\rm ACTPol}_{\rm sky}$ and $f^{\rm PLANCK}_{\rm sky}$ are fractal sky coverage
of ACTPol and PLANCK.


\end{document}